\documentclass[twocolumn,bibnotes,superscriptaddress,showpacs,amsmath,amssymb,longbibliography,aps,prx]{revtex4-2}

\usepackage{graphicx}
\usepackage{amsmath}
\usepackage{amssymb}
\usepackage{color}
\usepackage{comment}
\usepackage{bm}
\newcommand{\red}[1]{\textcolor{black}{#1}}
\newcommand{\blue}[1]{\textcolor{black}{#1}}
\usepackage[colorlinks=true,linkcolor=blue,citecolor=blue]{hyperref}

\begin{document}

\title{\red{Defect-Induced Low-Energy Majorana Excitations in the Kitaev Magnet $\alpha$-RuCl$_3$}}

\author{K.~Imamura}\email{imamura@qpm.k.u-tokyo.ac.jp}
\affiliation{Department of Advanced Materials Science, University of Tokyo, Kashiwa, Chiba 277-8561, Japan}
\author{Y.~Mizukami}
\thanks{Present adddress: Department of Physics, Tohoku University, Sendai 980-8578, Japan}
\author{O.~Tanaka}
\affiliation{Department of Advanced Materials Science, University of Tokyo, Kashiwa, Chiba 277-8561, Japan}
\author{R.~Grasset}
\author{M.~Konczykowski}
\affiliation{Laboratoire des Solides Irradi{\' e}s, CEA/DRF/IRAMIS, Ecole Polytechnique, CNRS, Institut Polytechnique de Paris, F-91128 Palaiseau, France}
\author{N.~Kurita}
\affiliation{Department of Physics, Tokyo Institute of Technology, Meguro, Tokyo 152-8551, Japan}
\author{H.~Tanaka}
\affiliation{Innovator and Inventor Development Platform, Tokyo Institute of Technology, Yokohama 226-8502, Japan}
\author{Y.~Matsuda}
\affiliation{Department of Physics, Kyoto University, Kyoto 606-8502, Japan}
\author{M.~G.~Yamada}
\affiliation{Department of Physics, Gakushuin University, Mejiro, Toshima-ku, Tokyo 171-8588, Japan}
\affiliation{Department of Physics, University of Tokyo, Bunkyo-ku, Tokyo 113-0033, Japan}
\author{K.~Hashimoto}
\author{T.~Shibauchi}\email{shibauchi@k.u-tokyo.ac.jp}
\affiliation{Department of Advanced Materials Science, University of Tokyo, Kashiwa, Chiba 277-8561, Japan}

\begin{abstract}
The excitations in the Kitaev spin liquid (KSL) can be described by Majorana fermions, which have characteristic field dependence of bulk gap and topological edge modes. In the high-field state of layered honeycomb magnet $\alpha$-RuCl$_3$, experimental results supporting these Majorana features have been reported recently. However, there are challenges due to sample dependence and the impact of inevitable disorder on the KSL is poorly understood. Here we study how low-energy excitations are modified by introducing point defects in $\alpha$-RuCl$_3$ using electron irradiation, which induces site vacancies and exchange randomness. High-resolution measurements of the temperature dependence of specific heat $C(T)$ under in-plane fields $H$ reveal that while the field-dependent Majorana gap is almost intact, additional low-energy states with $C/T=A(H)T$ are induced by introduced defects. At low temperatures, we obtain the data collapse of $C/T\sim H^{-\gamma}(T/H)$ expected for a disordered quantum spin system, but with an anomalously large exponent $\gamma$. This leads us to find \red{a power-law relationship between the coefficient $A(H)$ and} the field-sensitive Majorana gap. These results \red{are consistent with the picture} that the disorder induces low-energy linear Majorana excitations, which may be considered as a weak localization effect of Majorana fermions in the KSL. 
\end{abstract}

\maketitle

\section{Introduction}
In strongly interacting quantum systems, impurity effects are known to provide important information about their ground and excited states. For example, in anisotropic superconductors with a pairing mechanism different from conventional superconductivity, nonmagnetic impurities exert a strong pairing-breaking effect, which significantly affects the superconducting transition temperature, gap structure, and the density of states of quasiparticles. Among strongly interacting systems, quantum spin liquids (QSLs) in insulators have recently attracted much attention.  Although the spins in QSLs are strongly interacting with each other, they do not order even at absolute zero due to strong quantum fluctuations~\cite{Balents2010}. In the ground state of QSLs, strong quantum mechanical entanglement leads to the emergence of fractionalized exotic quasiparticles that are intimately related to the topology. However, the impact of randomness on the QSL states remains largely elusive.

An exactly solvable spin model that can give such QSLs is the Kitaev model of a two-dimensional honeycomb lattice~\cite{Kitaev2006}. 
In this model, the bond-dependent Ising interactions of magnitude $J$ introduce exchange frustrations, leading to a Kitaev spin liquid (KSL) ground state. An important property of this state is that the excitations can be described by two types of Majorana quasiparticles, namely, itinerant Majorana fermions and localized $Z_2$ fluxes (visons)~\cite{Kitaev2006}.

\begin{figure}[t]
    \includegraphics[width=1\linewidth]{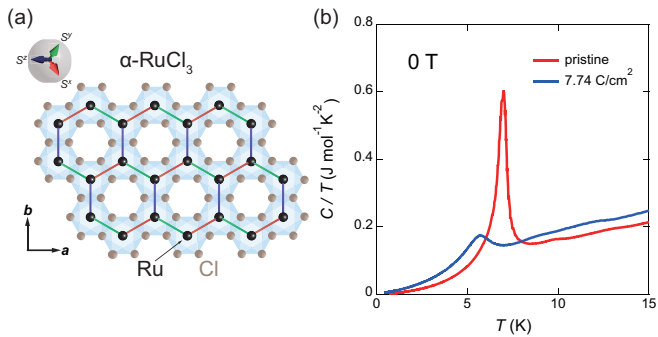}
    \caption{(a) Crystal structure of $\alpha$-RuCl$_3$ in the honeycomb plane. Ru$^{3+}$ ions (black circles) surrounded by octahedrons (shades) of Cl$^-$ ions (brown circles) form a layered honeycomb lattice. Upper inset shows the spin axis directions. (b) Temperature dependence of the specific heat divided by temperature $C/T$ for the pristine~\cite{Tanaka2022} and irradiated (with a dose of 7.74 C/cm$^2$) crystals of $\alpha$-RuCl$_3$. }
    \label{Fig1}
\end{figure}

The Kitaev interactions can be realized in real materials through the Jackeli-Khaliullin mechanism~\cite{Jackeli2009}, which boosted the search for candidates of Kitaev materials with honeycomb structures. The most studied is the spin-orbit assisted Mott insulator $\alpha$-RuCl$_3$~\cite{Takagi2019}, in which Ru$^{3+}$ ions are surrounded by octahedrons of Cl$^-$ ions, forming a layered honeycomb lattice (Fig.\,\ref{Fig1}(a)). The crystallographic $\bm{a}$ and $\bm{b}$ axes in the honeycomb plane are perpendicular and parallel to the Ru-Ru bond direction, respectively. Although it has been revealed that $\alpha$-RuCl$_3$ exhibits significant Kitaev interactions, it also has other magnetic interactions such as Heisenberg and off-diagonal terms~\cite{Suzuki2021}, which lead to an antiferromagnetic (AFM) order below the transition temperature $T_{\rm N}\sim7$\,K~\cite{Johnson2015}. However, this AFM order can be completely suppressed by applying a magnetic field of $\sim 8$\,T in the honeycomb plane~\cite{Yadav2016,Banerjee2018}. In the high-field quantum spin-disordered state, several anomalous features consistent with the expectations of the KSL have been observed experimentally. Most remarkably, the half-integer quantization of thermal Hall effect that occurs even in magnetic fields parallel to the plane (planar thermal Hall effect), together with the angle-dependent excitation gap, has been observed~\cite{Kasahara2018,Yokoi2021,Tanaka2022,Imamura2023}, which can be consistently explained by the topological bulk-edge correspondence of the Majorana fermions in the KSL.

On the other hand, it has been reported that the thermal Hall conductivity depends on the sample in terms of its magnitude and the temperature and field range over which it exhibits a plateau~\cite{Yamashita2020,Bruin2022,Kasahara2022,Czajka2023}. \red{These sample variations lead to the suggestions of alternative scenarios such as topological magnons~\cite{Czajka2023,McClarty2018,Joshi2018,Chern2021,Zhang2021} and phonons~\cite{lefranccois2022}. However, very recent observations of the gap closure for $\bm{H}\parallel\bm{b}$ argue against such bosonic scenarios~\cite{Imamura2023}. These active discussions on the nature of the high-field state of $\alpha$-RuCl$_3$ emphasize the crucial importance of elucidating the effects of disorder that inevitably exist in real materials.} Experimentally, the effect of site defects has been discussed in diluted $\alpha$-Ru$_{1-x}$Ir$_x$Cl$_3$~\cite{Lampen-Kelley2017,Do2018,Do2020}, which focuses on relatively high concentration regimes of defects. The bond disorder has also been discussed in the hydrogen intercalated iridate H$_3$LiIr$_2$O$_6$, which shows a divergent behavior in the low-temperature specific heat $C/T\propto T^{-1/2}$~\cite{Kitagawa2018}. Theoretical studies have pointed out that the introduction of site vacancies and bond disorder can give rise to various intriguing phenomena, such as a logarithmic growth of low-$T$ susceptibility, the divergent low-energy density of states, and the presence of local Majorana zero modes in the KSL~\cite{Willans2010,Willans2011,Knolle2019,Nasu2020,Sreenath2012,Petrova2013,Udagawa2018,Dantas2022,Singhania2023,Takahashi2022}. It is also notable that the disordered KSL shows localization of itinerant Majorana fermions, leading to a linear low-energy dispersion and a state called Anderson-Kitaev spin liquid with suppressed thermal Hall conductivity~\cite{Yamada2020,Kao2021}. However, the effects of disorder on the Majorana excitations in the high-field state of $\alpha$-RuCl$_3$ remain unclear.

\begin{figure*}[t]
    \includegraphics[width=0.85\linewidth]{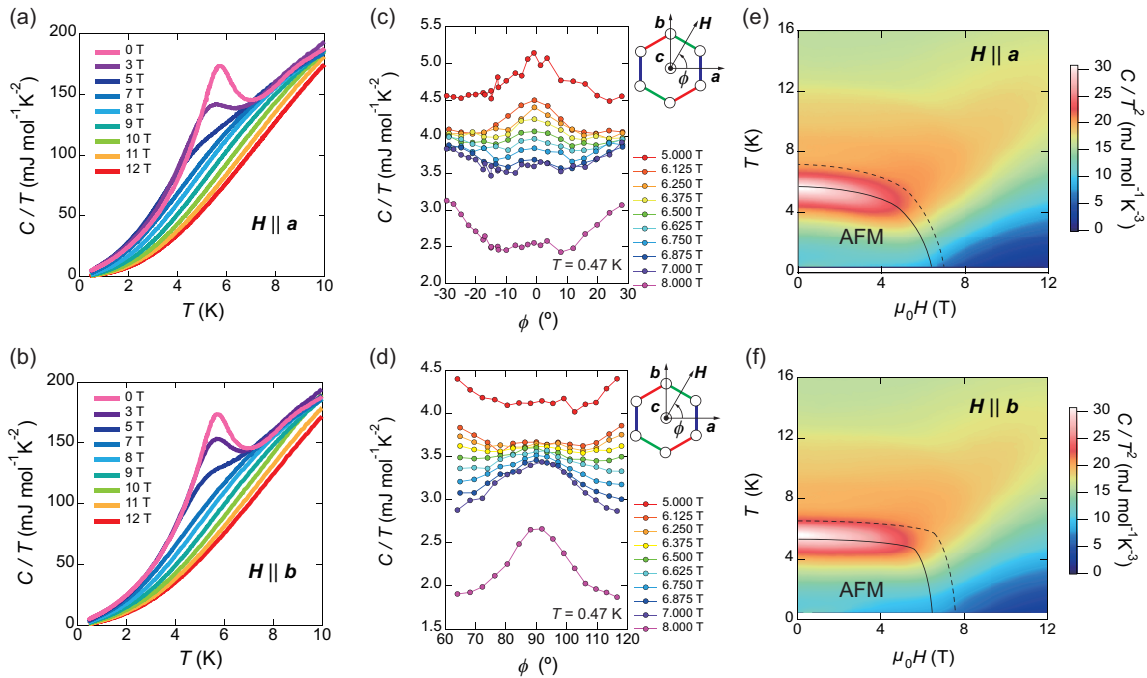}
    \caption{(a),(b) Temperature dependence of $C/T$ at several fields for $\bm{H}\parallel\bm{a}$ (a) and $\bm{H}\parallel\bm{b}$ (b) in irradiated $\alpha$-RuCl$_3$. (c),(d) In-plane field-angle dependence of $C/T$ at several fields at $T=0.47$ K around the $\bm{a}$ (c) and $\bm{b}$ (d) axes. (e),(f) $T$-$H$ phase diagrams of irradiated $\alpha$-RuCl$_3$ for $\bm{H}\parallel\bm{a}$ (e) and $\bm{H}\parallel\bm{b}$ (f). Superimposed color maps represent the magnitude of $C/T^2$. The solid (dashed) lines show the phase boundary between low-field antiferromagnetic and high-field KSL phases in the irradiated (pristine) sample.}
    \label{Fig2}
\end{figure*}

In this study, we use high-quality single crystals of $\alpha$-RuCl$_3$ and investigate the effects of point defects artificially introduced by electron irradiation. High-energy electron beams are used to create Frenkel pairs of vacancies and interstitial atoms, which act as point defects without changing the underlying electronic structure and lattice constants significantly~\cite{Mizukami2014,Cho2018}. We compare the low-energy excitations in pristine and irradiated crystals by high-resolution specific heat measurements down to $\sim 0.45$\,K under in-plane magnetic fields, which are recently recognized as a powerful probe of charge-neutral Majorana fermions~\cite{Tanaka2022,Imamura2023}. In the KSL, itinerant Majorana fermions exhibit gapless excitations at zero magnetic field, but the application of a magnetic field changes the low-energy gapless linear dispersion of Majorana fermions to a gapped one. When a magnetic field $\bm{H}=(h_x,h_y,h_z)$ is applied, the Majorana gap $\Delta_{\text M}$ opens with characteristic field dependence as $\Delta_{\text M}\propto|h_xh_yh_z|/\Delta_{\text{flux}}^2$, where $\Delta_{\text{flux}}$ represents the excitation gap of $Z_2$ fluxes~\cite{Kitaev2006}. Here $x$, $y$, and $z$ are the spin axes different from the crystallographic axes (Fig.\,\ref{Fig1}(a)). In the pristine sample, clear angle-dependent low-energy excitations fully consistent with the field-dependent Majorana gap have been observed~\cite{Tanaka2022,Imamura2023}. In irradiated $\alpha$-RuCl$_3$, we find a slight suppression of AFM order and the field-dependent Majorana gap similar to the pristine sample in the high-field state. In addition, we observe a field-dependent additional $T$-linear term in $C/T$, which indicates that low-energy excitations with a linear density of states emerge. At low temperatures, the \red{additional $T$-linear term in $C/T$ shows a peculiar relation} with Majorana excitation gap, which is consistent with recent theoretical calculations that the low-energy linear excitations can appear by the weak Anderson localization of Majorana fermions induced by disorder in $\alpha$-RuCl$_3$.

\section{Methods}
High-quality single crystals of $\alpha$-RuCl$_3$ were grown by the vertical Bridgman method~\cite{Kubota2015}. The lateral size of the sample used is $\sim1.1\times1.1$ mm$^2$ with a weight of $\sim340\,\mu$g. We employed the long-relaxation method, where a Cernox-1030 resistor serves as thermometer, heater, and sample stage. This experimental setup enables us to perform high-resolution heat capacity measurements under in-plane magnetic field rotation~\cite{Tanaka2022}. 

The electron irradiation with an incident electron energy of 2.5\,MeV was performed at 22\,K. The vacancies of Ru atoms created by the irradiation act as site vacancies of the honeycomb network, while defects in the surrounding Cl atoms (Fig.\,\ref{Fig1}(a)) are considered to act as bond disorder because the magnitude of Kitaev interactions is determined by the interorbital hopping mediated by Cl$^-$ ions~\cite{Jackeli2009}. In the present study, we use an irradiated crystal with a dose of 7.74\,C/cm$^2$, in which the concentrations of site vacancies and bond disorder in the Ru honeycomb network are both estimated as $\sim0.6$\% (see Appendix\,A).

\section{Results: Phase diagrams}
First, we discuss the change in the phase diagram by electron irradiation. Figure\,\ref{Fig1}(b) shows the temperature dependence of the specific heat divided by temperature $C/T$ for the pristine and irradiated samples. For the pristine sample, the sharp peak anomaly observed in $C/T$ at $T_{\rm N}\approx7$\,K represents the AFM transition~\cite{Tanaka2022}. For the irradiated sample, $T_N$ is shifted to a lower temperature $\sim5.7$\,K, and the peak in $C/T$ is suppressed. This behavior is consistent with the reported results in diluted $\alpha$-Ru$_{1-x}$Ir$_x$Cl$_3$~\cite{Do2018,Do2020}, \red{indicating that the long-range AFM order is vulnerable to disorder.} In $\alpha$-RuCl$_3$, stacking faults can be formed, and the associated $C/T$ anomalies appear at $T_{\rm N2}$ between $\sim10$ and 14\,K~\cite{Kubota2015,Cao2016}. However, these anomalies are barely visible, indicating no significant stacking faults in our samples.

Figures\,\ref{Fig2}(a) and (b) represent the temperature dependence of $C/T$ under several fields for $\bm{H}\parallel\bm{a}$ ($\parallel [11\bar{2}]$ in the spin coordinate) and $\bm{H}\parallel{\bm{b}}$ ($\parallel [\bar{1}10]$), respectively. As the in-plane magnetic field increases, $T_{\rm N}$ shifts to lower temperatures, and the anomalies in $C/T$ at $T_{\rm N}$ are getting smeared. To determine the critical fields at low temperatures, we use the field-angle rotation within the honeycomb plane, because the previous study in the pristine sample has shown that maxima and minima angles are interchanged between the low-field AFM and high-field disordered phases~\cite{Tanaka2022}. As shown in Figs.\,\ref{Fig2}(c) and (d), the field-angle dependence of $C/T$ at 0.47\,K changes from a peak to a dip structure near the $\bm{a}$ direction and vice versa near the $\bm{b}$ direction with increasing field. We define the critical field $H_{\rm c}$ as the magnetic field at which the angle dependence switches. For $\bm{H}\parallel\bm{a}$ ($\bm{b}$), $\mu_0H_{\rm c}$ is about 6.5\,T (6.7\,T). The obtained $T$-$H$ phase diagrams for the two field directions are shown in Figs.\,\ref{Fig2}(e) and (f). The solid line represents the phase boundary between the AFM and the high-field disordered phase. The boundary is shifted to lower temperatures and fields by irradiation.

\begin{figure*}[t]
    \includegraphics[width=1\linewidth]{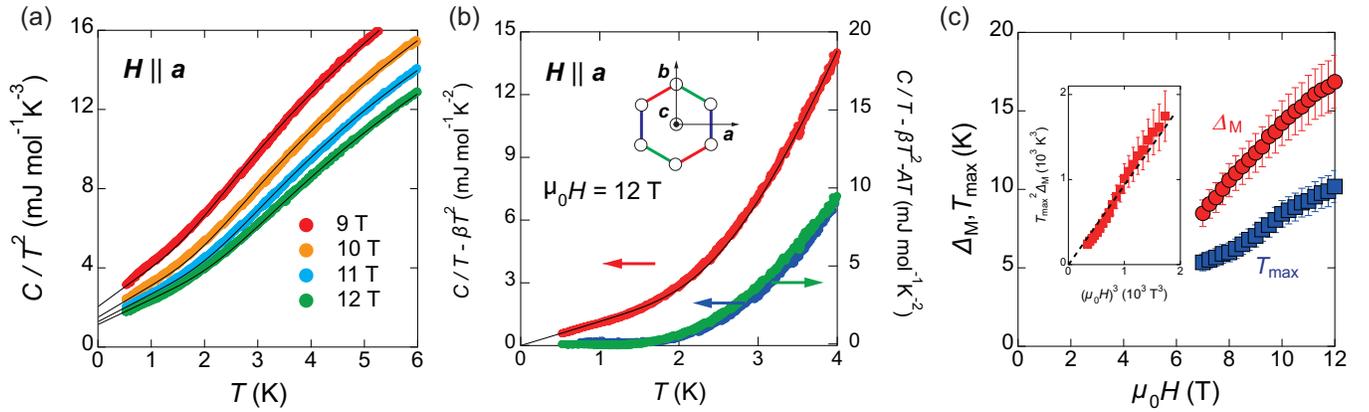}
    \caption{(a) Temperature dependence of $C/T^2$ for $\bm{H}\parallel\bm{a}$ at several fields in irradiated $\alpha$-RuCl$_3$. Solid lines represent the fitting curves with the excitation gap $\Delta_{\rm M}$ and the additional term $C/T^2=A(H)$. (b) Temperature dependence of $C/T-\beta T^2$ with $\beta=1.27$\,mJ\,mol$^{-1}$K$^{-4}$ for $\bm{H}\parallel\bm{a}$ at 12\,T (red) in irradiated $\alpha$-RuCl$_3$, compared with the previous data in pristine $\alpha$-RuCl$_3$ (blue)~\cite{Tanaka2022}. The solid line represents the same fitting curve as in (a). We also plot $C/T-\beta T^2-AT$ (green, right axis) in the irradiated sample. (c) Field dependence of Majorana gap $\Delta_{\rm M}$ (red circles) and $T_{\rm max}$ (blue squares) for $\bm{H}\parallel\bm{a}$ in irradiated $\alpha$-RuCl$_3$. The inset shows $T_{\rm max}^2\Delta_{\rm M}$ as a function of $(\mu_0H)^3$. The dashed line represents the fitting curve for the expected $H^3$ dependence. }
    \label{Fig3}
  \end{figure*}
  
\section{Results: Low-energy excitations}

Next, we discuss the low-energy excitations from the low-temperature behaviors of specific heat. In Fig.\,\ref{Fig3}(a), we plot the temperature dependence of $C/T^2$ for $\bm{H}\parallel\bm{a}$ in the high-field phase of irradiated $\alpha$-RuCl$_3$. It is clear from the data that a finite residual term $C/T^2$ exists in the zero-temperature limit. It has been established that the phonon contribution gives $\beta T^3$ term in $C(T)$ with $\beta\approx1.3$\,mJ\,mol$^{-1}$K$^{-4}$~\cite{Tanaka2022,Widmann2019}, which does not contribute to the residual $C/T^2$ term. Figure\,\ref{Fig3}(b) shows the $C/T-\beta T^2$ data for the irradiated sample, which display a $T$-linear behavior at low temperatures, corresponding to the finite residual $C/T^2$ term. In contrast, the data in the pristine sample show exponential temperature dependence, indicating fully gapped excitations~\cite{Tanaka2022}. When we subtract the $T$-linear term from $C/T-\beta T^2$ for the irradiated sample, we essentially obtain a gapped behavior similar to the pristine case (Fig.\,\ref{Fig3}(b)), although the prefactors that determine the magnitude are different possibly due to a slight change in the interactions. From these results we conclude that introduced disorder by irradiation induces additional low-energy density of states of quasiparticles.

\begin{figure}[t]
    \includegraphics[width=0.7\linewidth]{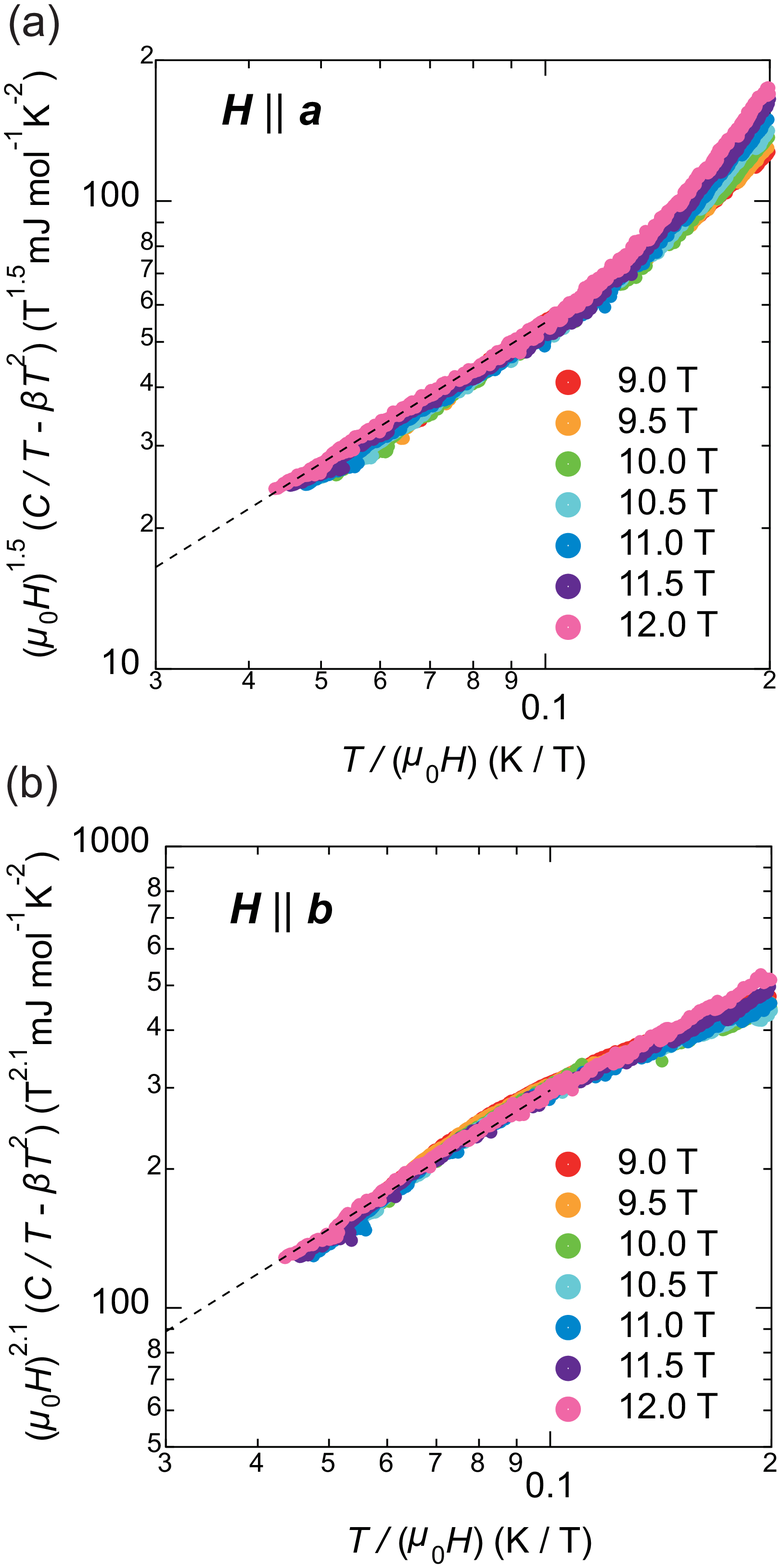}
    \caption{\red{Data collapse in irradiated $\alpha$-RuCl$_3$. $H^\gamma(C/T-\beta_{\rm ph}T^2)$ is plotted} versus $T/H$ with $\gamma=1.5$ for $\bm{H}\parallel\bm{a}$ (a) and $\gamma=2.1$ for $\bm{H}\parallel\bm{b}$ (b). Dashed lines represent $T/H$-linear dependence showing the $(T/H)^q$ data collapse with $q=1$.}
    \label{Fig4}
\end{figure}

To analyze the data more quantitatively, we employ the following formula: $C(T,H)/T=\beta(H)T^2+C_{\rm M}(T,H)/T+C_{\rm flux}(T,H)/T+A(H)T$. Here we consider bosonic contributions in the first $\beta T^2$ term, and the contributions from itinerant Majorana fermions ($C_{\rm M}/T$) and $Z_2$ fluxes ($C_{\rm flux}/T$) in the second and third terms, respectively. These three terms have been used in the previous study for the pristine sample~\cite{Tanaka2022}. Here, for the irradiated sample, we consider the fourth term $A(H)T$, which describes the observed additional low-energy excitations. The details of the fitting procedures are explained in Appendix\,B. We obtain fairly good fitting results as shown in Figs.\,\ref{Fig3}(a) and (b) by the solid lines, and the obtained fitting parameter $\Delta_{\text M}$ is about 16.7\,K for 12\,T, which is close to the value (14.4\,K) in the pristine sample~\cite{Tanaka2022}. 

Figure\,\ref{Fig3}(c) demonstrates the field dependence of the Majorana gap $\Delta_{\rm M}$ for $\bm{H}\parallel\bm{a}$ obtained from the fitting. The Majorana gap $\Delta_{\text M}$ shows a systematic increase as a function of field, which is in agreement with the previous reports for the pristine $\alpha$-RuCl$_3$~\cite{Tanaka2022,Sears2017,Wolter2017}. It has also been shown that the specific heat $C$, excluding the bosonic contribution $\beta T^3$, displays a broad peak at the peak temperature $T_{\text{max}}$ (see Appendix\,B), which is associated with the excitations of $Z_2$ fluxes~\cite{Motome2020,Do2017}. This $Z_2$ flux peak temperature is related to the Kitaev interaction $J$ and quantum Monte Carlo simulations suggest the relations $T_{\text{max}}\sim0.012J$ and $\Delta_{\text{flux}}\sim0.07J$~\cite{Motome2020}. Therefore, the simple relation $T_{\text{max}}\propto \Delta_{\text{flux}}$ can be used to verify the validity of the equation $\Delta_{\text M}\propto|h_xh_yh_z|/\Delta_{\text{flux}}^2$. In the pristine sample, the behavior of $T_{\text{max}}^2\Delta_{\text M}$ exhibits distinct $H^3$ dependence, which is well consistent with the Kitaev model~\cite{Tanaka2022}. From the combined results of $\Delta_{\text M}(H)$ and $T_{\text{max}}(H)$, we also find similar $H^3$ dependence of $T_{\text{max}}^2\Delta_{\text M}$ in the irradiated sample (Fig.\,\ref{Fig3}(c), inset). This finding provides strong evidence that the KSL is sustained even with $\sim0.6$\% defects in the high-field phase of $\alpha$-RuCl$_3$. 
\red{We note that the slope of the $T_{\text{max}}^2\Delta_{\text M}$ vs.\ $H^3$ plot is consistent with the pristine data within $\sim 10$\%, implying that the Majorana gap is almost intact after irradiation.}
\blue{The slight slope change near $\sim~10$\,T is associated with the slope changes in $\Delta_{\text M}(H)$ and $T_{\text{max}}(H)$, which are also seen in the pristine sample~\cite{Tanaka2022}. These changes may be related to the possible transition to a high-field state where the quantum thermal Hall effect vanishes~\cite{Kasahara2022,Tanaka2022,Suetsugu2022}, which deserves further studies.} 

\section{Discussion}

Here we examine whether the observed specific heat behaviors originate from the intrinsic properties of the KSL with quenched disorder. It has been proposed that the low-temperature specific heat $C(H, T)$ in spin-disordered systems exhibits $T/H$ data collapse, showing universal scaling features~\cite{Kimchi2018}. In frustrated disordered quantum spin systems, $C/T$ follows $H^{-\gamma}(T/H)^q$ at low temperatures, where $\gamma$ is a non-universal exponent and $q$ is an integer index of either 0, 1, or 2. \red{We appy this analysis} for the low-temperature data of $C/T-\beta T^2$ in the high-field phase of irradiated $\alpha$-RuCl$_3$. As shown in Figs.\,\ref{Fig4}(a) and (b), the low-temperature data show the $q=1$ collapse behavior for $\bm{H}\parallel\bm{a}$ ($\bm{b}$) with the exponent $\gamma\approx 1.5$ (2.1). These results strongly indicate that the low-energy excitations are governed by the introduced disorder in this system. In the random-singlet distribution of frustrated spin-disordered systems, the non-universal exponent is expected to be in the range of $0\lesssim\gamma\lesssim1$~\cite{Kimchi2018}, as observed in layered triangular and kagome antiferromagnets~\cite{Murayama2020,Murayama2022}. In the honeycomb iridate H$_3$LiIr$_2$O$_6$ under magnetic fields, the scaling with $\gamma\sim1/2$ within this range has also been reported~\cite{Kitagawa2018}. However, here we find anomalously large $\gamma$ values beyond this range, suggesting a different nature of the disorder-induced states in $\alpha$-RuCl$_3$. 

\begin{figure}[t]
    \includegraphics[width=0.7\linewidth]{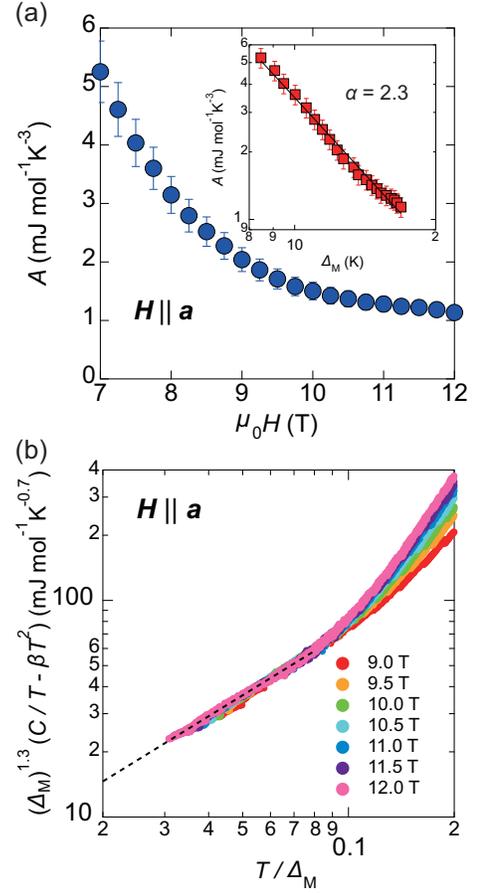}
    \caption{(a) Field dependence of the coefficient $A(H)$ of the additional $T$-linear term in $C/T$ for $\bm{H}\parallel\bm{a}$. The inset shows $A$ as a function of Majorana gap $\Delta_{\rm M}$. The solid line represents the fitting curve for the power-law \red{dependence} $A\propto \Delta_{\rm M}^{-\alpha}$ with the exponent $\alpha=2.3$. (b) \red{Plot} of $\Delta_{\rm M}^{(\alpha-1)}(C/T-\beta_{\rm ph}T^2)$ versus $T/\Delta_{\rm M}$ for $\bm{H}\parallel\bm{a}$.}
    \label{Fig5}
\end{figure}

The large $\gamma$ exponent immediately indicates that the additional $T$-linear behavior of $C/T$ observed in the irradiated sample with a small defect concentration has unusually strong dependence on magnetic field. The fact that the Majorana excitation gap also has strong sensitivity to magnetic field suggests that the defect-induced excitations are related to the Majorana physics in the KSL. 
Theoretical calculations for the gapped phase of the KSL under magnetic fields have shown that both bond disorder~\cite{Yamada2020} and site vacancies~\cite{Kao2021} can induce the in-gap localization modes of Majorana fermions, whose density of states shows a linear spectrum at low energies. 
Such in-gap modes suggested by these calculations can be compared with our data for $\bm{H}\parallel\bm{a}$, where the coefficient $A(H)$ in the induced $A(H)T$ term in $C/T$ inside the Majorana excitation gap $\Delta_{\rm M}$ can be evaluated quantitatively.

Figure\,\ref{Fig5}(a) displays the field dependence of the coefficient $A$ in the high-field state for $\bm{H}\parallel\bm{a}$. For the gapped KSL, it has been suggested that the in-gap energy-linear localization modes induced by the disorder has a relation with the three-spin interaction term $\kappa$ ($\sim|h_xh_yh_z|/J^2$)~\cite{Kao2021}, which has essentially the same form as Majorana gap $\Delta_{\rm M}$. \red{Here we find a peculiar power-law relationship between $A$ and} $\Delta_{\rm M}$ ($\propto\kappa$) over a wide range of magnetic fields; $A(H)\propto \Delta_{\rm M}(H)^{-\alpha}$ with an exponent $\alpha\approx2.3$ (Fig.\,\ref{Fig5}(a), inset). This is in very good agreement with the theoretical finding of the power-law \red{dependence} $C/T\sim \kappa^{-\alpha} T$ with $\alpha\sim2.25$ for the KSL with 2\% site vacancies~\cite{Kao2021}. This \red{power-law relationship between the disorder-induced $A(H)T$ term in $C/T$ and $\Delta_{\rm M}$} corresponds to the large exponent $\gamma\approx1.5$ for the data collapse analysis, which suggests $A\propto H^{-2.5}$. The fact that $T_{\rm max}\propto \Delta_{\rm flux}\propto J$ has some field dependence (Fig.\,\ref{Fig3}(c)) implies that the \red{relation} with $\Delta_{\rm M}$ rather than that with $H$ is more straightforward, and indeed a clear data collapse behavior is also seen at low temperatures in such a \red{$\Delta_{\rm M}$-based} analysis (Fig.\,\ref{Fig5}(b)). These \red{field-sensitive behaviors of the additional $A(H)T$ term strongly suggest} that the Majorana physics is at play in the disorder-induced low-energy excitations.

These results \red{are consistent with the picture} that the introduced defects induce low-energy linear Majorana excitations in the high-field KSL state of $\alpha$-RuCl$_3$. Such energy-linear excitations can be interpreted by the weak Anderson localization effect of Majorana fermions~\cite{Yamada2020,Kao2021}. Based on the symmetry classification, the KSL in the absence of a magnetic field belongs to the symmetry class called $BD$I, but when time-reversal symmetry is broken by the application of magnetic fields the symmetry class changes to $D$~\cite{Altland1997,O'Brien2016}. In class $D$, it is a general property of weak localization that the density of states exhibits a linear behavior at low energies in the presence of weak disorder~\cite{Ryu2001}. Therefore, the $T$-linear behavior of $C/T$ observed in the high-field state of irradiated $\alpha$-RuCl$_3$ is consistent with the weak Anderson localization picture in the symmetry class $D$. 

It is remarkable that with only $\sim0.6$\% defects a significant amount of the anomalous localization term in the specific heat appears at low energies, implying that the KSL is very sensitive to randomness. 
The disorder-induced weak localization of Majorana fermions is expected to influence the thermal Hall effect in the gapped KSL, which is closely related to the edge current of itinerant Majorana fermions. Recent calculations for the bond disorder have shown that the thermal Hall conductivity $\kappa_{xy}$ is suppressed by the localization effect~\cite{Yamada2020}. As the randomness is inevitably present even in the pristine samples, such disorder effects provide a source of sample variations of the $\kappa_{xy}$ values in the high-field state of $\alpha$-RuCl$_3$. Indeed, a recent thermal transport study reported a tendency that the samples with higher longitudinal thermal conductivity $\kappa_{xx}$ exhibit larger $\kappa_{xy}$ values closer to the half-integer quantization~\cite{Kasahara2022}, implying that the long mean free path of quasiparticles, which is sensitive to disorder, is an important requisite for the observation of Majorana quantization. 

In conclusion, we conducted high-resolution specific heat under in-plane magnetic fields in irradiated $\alpha$-RuCl$_3$. The introduction of disorder through electron irradiation resulted in the slight suppression of the antiferromagnetic order. The field dependence of the Majorana gap provides thermodynamic evidence that the KSL is maintained in the high-filed state of the disordered $\alpha$-RuCl$_3$. Importantly, we observed an additional in-gap $T$-linear term in $C/T$, whose coefficient $A(H)$ shows distinct \red{field-sensitive behaviors suggestive} of Majorana physics in the KSL. These results may be interpreted by the weak localization of Majorana fermions, which is induced by the disorder. Our experimental study points to the importance of the disorder effects on the low-energy Majorana excitations in the KSL. 

\section*{Acknowledgements}
We thank S.\ Fujimoto, E.-G.\ Moon, J.\ Nasu, and M.~O.\ Takahashi for fruitful discussions. The authors acknowledge support from the EMIR\&A French network (FR CNRS 3618) on the platform SIRIUS. 
This work was supported by CREST (No.\ JPMJCR19T5) from Japan Science and Technology (JST), Grants-in-Aid for Scientific Research (KAKENHI) (Nos.\ JP22H00105, JP22K18683, JP22K14005, JP21H01793, JP19H00649, and JP18H05227), and Grant-in-Aid for Scientific Research on Innovative Areas “Quantum Liquid Crystals” (No.\ JP19H05824) from Japan Society for the Promotion of Science. M.G.Y. is supported by PRESTO (No.\ JPMJPR225B) and by the Center of Innovations for Sustainable Quantum AI (No.\ JPMJPF2221) from JST.

\section*{Appendix}

\subsection{Electron irradiation} 

Electron irradiation was performed by using the SIRIUS Pelletron accelerator operated by the Laboratoire des Solides Irradi{\' e}s (LSI) at {\' E}cole Polytechnique, maintaining the sample immersed in liquid hydrogen at 22\,K during irradiation. The low-temperature environment is important to prevent defect migration and agglomeration. Partial annealing of the introduced defects occurs upon warming to room temperatures~\cite{Prozorov2014}. 
\red{The penetration depth of the 2.5-MeV electron beam is estimated as $\sim 4.6$\,mm, which is much longer than the sample thickness of $\sim100\,\mu$m. This ensures the uniform distribution of the point defects.}

\begin{figure}[b]
    \includegraphics[width=0.7\linewidth]{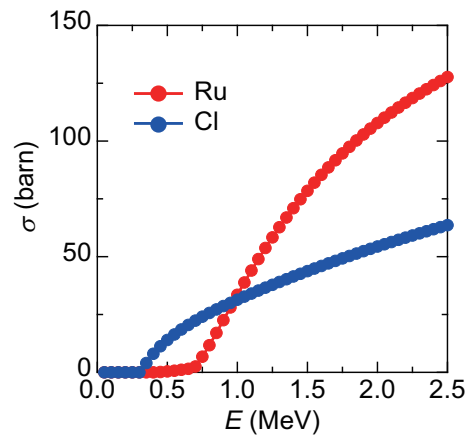}
    \caption{Calculated cross-section $\sigma$ for Frenkel-pair production at the Ru (red circles) and Cl (blue circles) sites of $\alpha$-RuCl$_3$ as a function of incident electron energy $E$.}
    \label{Fig6}
\end{figure}

We calculated the cross-sections $\sigma$ for collisions of high-energy electrons as a function of incident electron energy and estimated the amount of created defects (Frenkel pairs) on the Ru and Cl sites by using the SECTE software developed at LSI~\cite{Bois1987}. We use a typical value of $E_{\rm d}=25$\,eV for the energy of ion displacement~\cite{Roppongi2023}. In Fig.\,\ref{Fig6}, the cross-sections $\sigma$ of Ru and Cl sites are plotted as a function of incident energy $E$ of electrons. We find that $\sigma$ is twice larger for Ru than that for Cl for the present incident energy $E=2.5$\,MeV. The density of vacancies of Ru atoms, $v_{\rm Ru}$, can be estimated by $v_{\rm Ru}=\sigma \phi n$, where $\phi$ is the irradiation fluence (the number of high-energy electrons incident per unit area of the sample) and $n$ is the density of atoms per unit volume. \red{For the dose of 7.74\,C/cm$^2$, we} estimate $v_{\rm Ru}/n\sim 0.6$\%, which corresponds to the concentration of site vacancies in the honeycomb lattice. Similarly, the vacancy concentration for Cl atoms is estimated as $\sim 0.3$\%. As the exchange interaction for each bond is determined via two Cl$^{-}$ ions, the bond disorder concentration is also estimated as $\sim 0.6$\% of the total bonds.

\begin{table}[t]
    \red{\caption{\label{tab:lattice}Lattice constants in unit of \AA\ of $C2/m$ structure at room temperature.}}
     \begin{ruledtabular}
     \begin{tabular}{cccc}
     Crystal & $a$ (\AA) & $b$ (\AA) & $c$ (\AA)\\
     Pristine & 5.980&10.347&6.034\\
     Pristine~\cite{Bruin2022} & 6.041&10.416&6.088\\
     Irradiated (12.52\,C/cm$^2$) & 5.998&10.386&6.055\\  
 \end{tabular}
     \end{ruledtabular}
     \end{table}

 \begin{figure}[b]
     \includegraphics[width=0.6\linewidth]{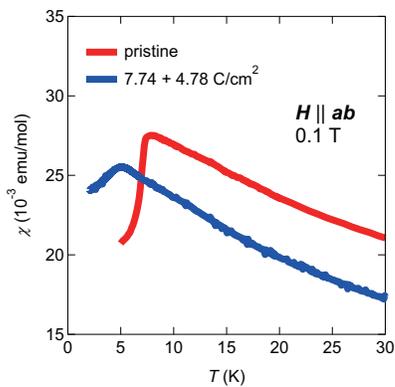}
     \red{\caption{
         Temperature dependence of magnetic susceptibility $\chi(T)$ measured for an in-plane field of 0.1\,T in the irradiated $\alpha$-RuCl$_3$ with a toal dose of 12.52\,C/cm$^2$ (blue) compared with the previous data for the pristine one (red)~\cite{Tanaka2022}.}
         }
     \label{fig:chi}
 \end{figure}
 
\red{After the specific heat measurements, we have made additional irradiation with the dose of 4.78\,C/cm$^2$ corresponding to the total vacancy density of $\sim 1$\% and we have performed X-ray diffraction and magnetic susceptibility measurements. The lattice constants for the $C2/m$ structure at room temperature are shown in Table\,\ref{tab:lattice}. For comparison, the lattice parameters measured for a pristine sample and reported values for a Bridgman sample~\cite{Bruin2022} are also shown. We find that these parameters are essentially unchanged after irradiation considering the sample-dependent uncertainties. The magnetic susceptibility $\chi(T)$ measured under in-plane fields of 0.1\,T shows a single transition at $\sim5$\,K, which is shifted down from the original $T_{\rm N}\sim7$\,K in the pristine sample, without showing higher transition anomalies associated with the stacking faults as shown in Fig.\,7. We also find no signature of magnetic impurities, indicating that the irradiation introduces nonmagnetic point defects. At higher temperatures, the hysteresis behavior in $\chi(T)$ associated with the structural transition from high-temperature $C2/m$ to low-temperature $R\bar{3}$~\cite{Nagai2020,Kim2023} is not observed for the irradiated sample, suggesting that the low-temperature structure remains $C2/m$. This is consistent with the two-fold rotational symmetry as found in the slight difference of $C/T$ at $\phi=\pm30^\circ$ and $\phi=90^\circ$. Possible effects of this two-fold symmetry need to be clarified by further studies, but in our previous study~\cite{Tanaka2022} we found that the rotational symmetry breaking at high fields where the Majorana gap behavior remains essentially unchanged. Together with the fact that the Majorana gap behavior in Fig.\,\ref{Fig3}(b) is almost intact from the pristine sample when we subtract the induced $T$-linear term, these results strongly suggest that the additional low-energy excitations are induced by the introduced defects rather than the lowered symmetry.
}

\subsection{Fitting procedures} 

The contribution of itinerant Majorana fermions $C_{\text{M}}$ to the specific heat can be evaluated from the standard calculations assuming that quasiparticles form a gas of fermions;
\begin{eqnarray}
C_{\text{M}}(T;\Delta_{\text{M}})&=&\frac{T}{V}\frac{\partial S(T)}{\partial T}  \nonumber\\
&=&\frac{1}{V}\sum_k\left(\frac{E(k)}{T}\right)^2n_{\text{F}}(E(k))(1-n_{\text{F}}(E(k))) \nonumber
\end{eqnarray}
with the Fermi-Dirac distribution function $n_{\text{F}}(x)=(\text{e}^x+1)^{-1}$~\cite{Tanaka2022}. We use the two-dimensional model for energy dispersion $E=\sqrt{v^2|{\bm k}|^2+\Delta_{\text{M}}^2}$, and the low-temperature specific heat is given by
\begin{equation}
C_{\text{M}}(T;\Delta_{\text{M}})/T=\frac{\mathcal G(\infty)}{v^2}\left(1-\frac{\mathcal G(\Delta_{\text{M}}/T)}{\mathcal G(\infty)}\right). \nonumber
\end{equation}
Here, the one-parameter function $\mathcal G(y)$ is given by $\mathcal G(y)=\int_0^y\frac{dx}{2\pi}\frac{x^3\text{e}^x}{(\text{e}^x+1)^2}$. For $\bm{H}\parallel\bm{b}$ the Majorana gap is zero ($\Delta_{\text{M}}=0$), and $C_{\text{M}}/T$ takes the form $\alpha T$ with $\alpha=\mathcal G(\infty)/v^2$, corresponding to a gapless linear dispersion. In the pristine sample, $C/T^2$ shows finite residual values, indicating the presence of $\alpha T$ terms in $C/T$~\cite{Tanaka2022}. 

\begin{figure}[t]
    \includegraphics[width=1.0\linewidth]{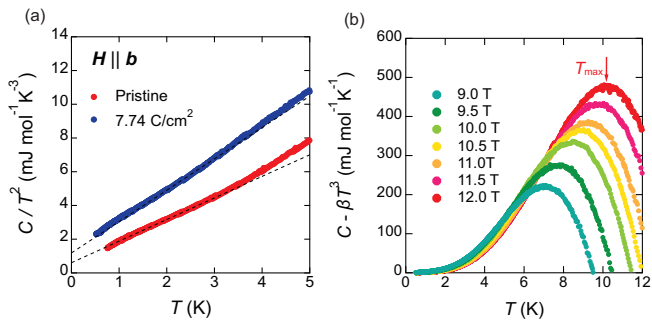}
    \caption{(a) Temperature dependence of $C/T^2$ for $\bm{H}\parallel\bm{b}$ at 12\,T in the irradiated $\alpha$-RuCl$_3$ (blue) compared with the previous data for the pristine one (red)~\cite{Tanaka2022}. The dashed lines represent the linear fitting curves. (b) Temperature dependence of $C-\beta T^3$ for $\bm{H}\parallel\bm{a}$, from which we determine $T_{\rm max}$ (arrow).}
    \label{Fig8}
\end{figure}

First, we focus on the $\bm{H}\parallel\bm{b}$ data. Figure\,\ref{Fig8}(a) shows the temperature dependence of $C/T^2$ at 12\,T in the pristine and irradiated samples. The $C/T^2$ data exhibit nearly $T$-linear behavior with finite residual values in the low-temperature limit, which can be fitted with $\alpha+\beta T$. The residual term $\alpha$ of the irradiated sample is about twice larger than that of the pristine sample, indicating that the additional excitation also appears for $\bm{H}\parallel\bm{b}$.

We consider the derivation from the $T$-linear behavior at high temperatures by the $Z_2$-flux contribution. The $Z_2$-flux contribution can be estimated by the simple Schottky formula $C_{\text{flux}}/T=A_{\text{flux}}\left(\Delta_{\text{flux}}^2/T^3\right) \cdot \text{e}^{\Delta_{\text{flux}}/T}\left(1+\text{e}^{\Delta_{\text{flux}}/T}\right)^2$. This contribution is weighted largely at high temperatures, because the $Z_2$-flux gap $\Delta_{\text{flux}}$, which is related to $T_{\rm max}$, is much larger than the Majorana gap $\Delta_{\text{M}}$ (Fig.\,\ref{Fig8}(b)). Therefore, we perform our fitting at low temperatures below $\frac{2}{3}T_{\text{max}}$, where the $Z_2$-flux contribution is small. To analyze the data for $\bm{H}\parallel\bm{a}$, we assume that the $Z_2$-flux term remains the same as that for $\bm{H}\parallel\bm{b}$, because $T_{\rm max}$ is almost angle independent~\cite{Tanaka2022}. By using these fixed parameters, we fit the $C(H,T)/T$ data with remaining variable parameters in the form $\beta(H)T^2+C_{\text{M}}(T;\Delta_{\text{M}})/T+A(H)T$.

\bibliography{refRuCl3_irradiate.bib}

\begin{thebibliography}{58}%
\makeatletter
\providecommand \@ifxundefined [1]{%
 \@ifx{#1\undefined}
}%
\providecommand \@ifnum [1]{%
 \ifnum #1\expandafter \@firstoftwo
 \else \expandafter \@secondoftwo
 \fi
}%
\providecommand \@ifx [1]{%
 \ifx #1\expandafter \@firstoftwo
 \else \expandafter \@secondoftwo
 \fi
}%
\providecommand \natexlab [1]{#1}%
\providecommand \enquote  [1]{``#1''}%
\providecommand \bibnamefont  [1]{#1}%
\providecommand \bibfnamefont [1]{#1}%
\providecommand \citenamefont [1]{#1}%
\providecommand \href@noop [0]{\@secondoftwo}%
\providecommand \href [0]{\begingroup \@sanitize@url \@href}%
\providecommand \@href[1]{\@@startlink{#1}\@@href}%
\providecommand \@@href[1]{\endgroup#1\@@endlink}%
\providecommand \@sanitize@url [0]{\catcode `\\12\catcode `\$12\catcode `\&12\catcode `\#12\catcode `\^12\catcode `\_12\catcode `\%12\relax}%
\providecommand \@@startlink[1]{}%
\providecommand \@@endlink[0]{}%
\providecommand \url  [0]{\begingroup\@sanitize@url \@url }%
\providecommand \@url [1]{\endgroup\@href {#1}{\urlprefix }}%
\providecommand \urlprefix  [0]{URL }%
\providecommand \Eprint [0]{\href }%
\providecommand \doibase [0]{https://doi.org/}%
\providecommand \selectlanguage [0]{\@gobble}%
\providecommand \bibinfo  [0]{\@secondoftwo}%
\providecommand \bibfield  [0]{\@secondoftwo}%
\providecommand \translation [1]{[#1]}%
\providecommand \BibitemOpen [0]{}%
\providecommand \bibitemStop [0]{}%
\providecommand \bibitemNoStop [0]{.\EOS\space}%
\providecommand \EOS [0]{\spacefactor3000\relax}%
\providecommand \BibitemShut  [1]{\csname bibitem#1\endcsname}%
\let\auto@bib@innerbib\@empty
\bibitem [{\citenamefont {Balents}(2010)}]{Balents2010}%
  \BibitemOpen
  \bibfield  {author} {\bibinfo {author} {\bibfnamefont {L.}~\bibnamefont {Balents}},\ }\bibfield  {title} {\bibinfo {title} {Spin liquids in frustrated magnets},\ }\href {https://doi.org/10.1038/nature08917} {\bibfield  {journal} {\bibinfo  {journal} {{Nature}}\ }\textbf {\bibinfo {volume} {464}},\ \bibinfo {pages} {199} (\bibinfo {year} {2010})}\BibitemShut {NoStop}%
\bibitem [{\citenamefont {{Kitaev}}(2006)}]{Kitaev2006}%
  \BibitemOpen
  \bibfield  {author} {\bibinfo {author} {\bibfnamefont {A.}~\bibnamefont {{Kitaev}}},\ }\bibfield  {title} {\bibinfo {title} {Anyons in an exactly solved model and beyond},\ }\href {https://doi.org/10.1016/j.aop.2005.10.005} {\bibfield  {journal} {\bibinfo  {journal} {Ann. Phys. (N. Y.)}\ }\textbf {\bibinfo {volume} {321}},\ \bibinfo {pages} {2} (\bibinfo {year} {2006})}\BibitemShut {NoStop}%
\bibitem [{\citenamefont {Jackeli}\ and\ \citenamefont {Khaliullin}(2009)}]{Jackeli2009}%
  \BibitemOpen
  \bibfield  {author} {\bibinfo {author} {\bibfnamefont {G.}~\bibnamefont {Jackeli}}\ and\ \bibinfo {author} {\bibfnamefont {G.}~\bibnamefont {Khaliullin}},\ }\bibfield  {title} {\bibinfo {title} {{Mott Insulators in the Strong Spin-Orbit Coupling Limit: From Heisenberg to a Quantum Compass and Kitaev Models}},\ }\href {https://doi.org/10.1103/PhysRevLett.102.017205} {\bibfield  {journal} {\bibinfo  {journal} {Phys. Rev. Lett.}\ }\textbf {\bibinfo {volume} {102}},\ \bibinfo {pages} {017205} (\bibinfo {year} {2009})}\BibitemShut {NoStop}%
\bibitem [{\citenamefont {Takagi}\ \emph {et~al.}(2019)\citenamefont {Takagi}, \citenamefont {Takayama}, \citenamefont {Jackeli}, \citenamefont {Khaliullin},\ and\ \citenamefont {Nagler}}]{Takagi2019}%
  \BibitemOpen
  \bibfield  {author} {\bibinfo {author} {\bibfnamefont {H.}~\bibnamefont {Takagi}}, \bibinfo {author} {\bibfnamefont {T.}~\bibnamefont {Takayama}}, \bibinfo {author} {\bibfnamefont {G.}~\bibnamefont {Jackeli}}, \bibinfo {author} {\bibfnamefont {G.}~\bibnamefont {Khaliullin}},\ and\ \bibinfo {author} {\bibfnamefont {S.~E.}\ \bibnamefont {Nagler}},\ }\bibfield  {title} {\bibinfo {title} {Concept and realization of {Kitaev} quantum spin liquids},\ }\href {https://doi.org/10.1038/s42254-019-0038-2} {\bibfield  {journal} {\bibinfo  {journal} {Nat. Rev. Phys.}\ }\textbf {\bibinfo {volume} {1}},\ \bibinfo {pages} {264} (\bibinfo {year} {2019})}\BibitemShut {NoStop}%
\bibitem [{\citenamefont {Suzuki}\ \emph {et~al.}(2021)\citenamefont {Suzuki}, \citenamefont {Liu}, \citenamefont {Bertinshaw}, \citenamefont {Ueda}, \citenamefont {Kim}, \citenamefont {Laha}, \citenamefont {Weber}, \citenamefont {Yang}, \citenamefont {Wang}, \citenamefont {Takahashi} \emph {et~al.}}]{Suzuki2021}%
  \BibitemOpen
  \bibfield  {author} {\bibinfo {author} {\bibfnamefont {H.}~\bibnamefont {Suzuki}}, \bibinfo {author} {\bibfnamefont {H.}~\bibnamefont {Liu}}, \bibinfo {author} {\bibfnamefont {J.}~\bibnamefont {Bertinshaw}}, \bibinfo {author} {\bibfnamefont {K.}~\bibnamefont {Ueda}}, \bibinfo {author} {\bibfnamefont {H.}~\bibnamefont {Kim}}, \bibinfo {author} {\bibfnamefont {S.}~\bibnamefont {Laha}}, \bibinfo {author} {\bibfnamefont {D.}~\bibnamefont {Weber}}, \bibinfo {author} {\bibfnamefont {Z.}~\bibnamefont {Yang}}, \bibinfo {author} {\bibfnamefont {L.}~\bibnamefont {Wang}}, \bibinfo {author} {\bibfnamefont {H.}~\bibnamefont {Takahashi}}, \emph {et~al.},\ }\bibfield  {title} {\bibinfo {title} {{Proximate ferromagnetic state in the Kitaev model material $\alpha$-RuCl$_3$}},\ }\href {https://doi.org/https://doi.org/10.1038/s41467-021-24722-4} {\bibfield  {journal} {\bibinfo  {journal} {Nat. Commun.}\ }\textbf {\bibinfo {volume} {12}},\ \bibinfo {pages} {4512} (\bibinfo {year} {2021})}\BibitemShut {NoStop}%
\bibitem [{\citenamefont {Johnson}\ \emph {et~al.}(2015)\citenamefont {Johnson}, \citenamefont {Williams}, \citenamefont {Haghighirad}, \citenamefont {Singleton}, \citenamefont {Zapf}, \citenamefont {Manuel}, \citenamefont {Mazin}, \citenamefont {Li}, \citenamefont {Jeschke}, \citenamefont {Valent\'{\i}},\ and\ \citenamefont {Coldea}}]{Johnson2015}%
  \BibitemOpen
  \bibfield  {author} {\bibinfo {author} {\bibfnamefont {R.~D.}\ \bibnamefont {Johnson}}, \bibinfo {author} {\bibfnamefont {S.~C.}\ \bibnamefont {Williams}}, \bibinfo {author} {\bibfnamefont {A.~A.}\ \bibnamefont {Haghighirad}}, \bibinfo {author} {\bibfnamefont {J.}~\bibnamefont {Singleton}}, \bibinfo {author} {\bibfnamefont {V.}~\bibnamefont {Zapf}}, \bibinfo {author} {\bibfnamefont {P.}~\bibnamefont {Manuel}}, \bibinfo {author} {\bibfnamefont {I.~I.}\ \bibnamefont {Mazin}}, \bibinfo {author} {\bibfnamefont {Y.}~\bibnamefont {Li}}, \bibinfo {author} {\bibfnamefont {H.~O.}\ \bibnamefont {Jeschke}}, \bibinfo {author} {\bibfnamefont {R.}~\bibnamefont {Valent\'{\i}}},\ and\ \bibinfo {author} {\bibfnamefont {R.}~\bibnamefont {Coldea}},\ }\bibfield  {title} {\bibinfo {title} {Monoclinic crystal structure of $\alpha$-{RuCl}$_{3}$ and the zigzag antiferromagnetic ground state},\ }\href {https://doi.org/10.1103/PhysRevB.92.235119} {\bibfield  {journal} {\bibinfo  {journal} {Phys. Rev. B}\ }\textbf {\bibinfo {volume} {92}},\ \bibinfo {pages} {235119} (\bibinfo {year} {2015})}\BibitemShut {NoStop}%
\bibitem [{\citenamefont {Yadav}\ \emph {et~al.}(2016)\citenamefont {Yadav}, \citenamefont {Bogdanov}, \citenamefont {Katukuri}, \citenamefont {Nishimoto}, \citenamefont {van~den Brink},\ and\ \citenamefont {Hozoi}}]{Yadav2016}%
  \BibitemOpen
  \bibfield  {author} {\bibinfo {author} {\bibfnamefont {R.}~\bibnamefont {Yadav}}, \bibinfo {author} {\bibfnamefont {N.~A.}\ \bibnamefont {Bogdanov}}, \bibinfo {author} {\bibfnamefont {V.~M.}\ \bibnamefont {Katukuri}}, \bibinfo {author} {\bibfnamefont {S.}~\bibnamefont {Nishimoto}}, \bibinfo {author} {\bibfnamefont {J.}~\bibnamefont {van~den Brink}},\ and\ \bibinfo {author} {\bibfnamefont {L.}~\bibnamefont {Hozoi}},\ }\bibfield  {title} {\bibinfo {title} {{Kitaev} exchange and field-induced quantum spin-liquid states in honeycomb $\alpha$-{RuCl}$_{3}$},\ }\href {https://doi.org/10.1038/srep37925} {\bibfield  {journal} {\bibinfo  {journal} {Sci. Rep.}\ }\textbf {\bibinfo {volume} {6}},\ \bibinfo {pages} {37925} (\bibinfo {year} {2016})}\BibitemShut {NoStop}%
\bibitem [{\citenamefont {Banerjee}\ \emph {et~al.}(2018)\citenamefont {Banerjee}, \citenamefont {Lampen-Kelley}, \citenamefont {Knolle}, \citenamefont {Balz}, \citenamefont {Aczel}, \citenamefont {Winn}, \citenamefont {Liu}, \citenamefont {Pajerowski}, \citenamefont {Yan}, \citenamefont {Bridges}, \citenamefont {Savici}, \citenamefont {Chakoumakos}, \citenamefont {Lumsden}, \citenamefont {Tennant}, \citenamefont {Moessner}, \citenamefont {Mandrus},\ and\ \citenamefont {Nagler}}]{Banerjee2018}%
  \BibitemOpen
  \bibfield  {author} {\bibinfo {author} {\bibfnamefont {A.}~\bibnamefont {Banerjee}}, \bibinfo {author} {\bibfnamefont {P.}~\bibnamefont {Lampen-Kelley}}, \bibinfo {author} {\bibfnamefont {J.}~\bibnamefont {Knolle}}, \bibinfo {author} {\bibfnamefont {C.}~\bibnamefont {Balz}}, \bibinfo {author} {\bibfnamefont {A.~A.}\ \bibnamefont {Aczel}}, \bibinfo {author} {\bibfnamefont {B.}~\bibnamefont {Winn}}, \bibinfo {author} {\bibfnamefont {Y.}~\bibnamefont {Liu}}, \bibinfo {author} {\bibfnamefont {D.}~\bibnamefont {Pajerowski}}, \bibinfo {author} {\bibfnamefont {J.}~\bibnamefont {Yan}}, \bibinfo {author} {\bibfnamefont {C.~A.}\ \bibnamefont {Bridges}}, \bibinfo {author} {\bibfnamefont {A.~T.}\ \bibnamefont {Savici}}, \bibinfo {author} {\bibfnamefont {B.~C.}\ \bibnamefont {Chakoumakos}}, \bibinfo {author} {\bibfnamefont {M.~D.}\ \bibnamefont {Lumsden}}, \bibinfo {author} {\bibfnamefont {D.~A.}\ \bibnamefont {Tennant}}, \bibinfo {author} {\bibfnamefont {R.}~\bibnamefont {Moessner}}, \bibinfo {author} {\bibfnamefont {D.~G.}\ \bibnamefont {Mandrus}},\ and\ \bibinfo {author} {\bibfnamefont {S.~E.}\ \bibnamefont {Nagler}},\ }\bibfield  {title} {\bibinfo {title} {Excitations in the field-induced quantum spin liquid state of $\alpha$-{RuCl}$_{3}$},\ }\href {https://doi.org/10.1038/s41535-018-0079-2} {\bibfield  {journal} {\bibinfo  {journal} {npj Quantum Mater.}\ }\textbf {\bibinfo {volume} {3}},\ \bibinfo {pages} {8} (\bibinfo {year} {2018})}\BibitemShut {NoStop}%
\bibitem [{\citenamefont {Kasahara}\ \emph {et~al.}(2018)\citenamefont {Kasahara}, \citenamefont {Ohnishi}, \citenamefont {Mizukami}, \citenamefont {Tanaka}, \citenamefont {Ma}, \citenamefont {Sugii}, \citenamefont {Kurita}, \citenamefont {Tanaka}, \citenamefont {Nasu}, \citenamefont {Motome}, \citenamefont {Shibauchi},\ and\ \citenamefont {Matsuda}}]{Kasahara2018}%
  \BibitemOpen
  \bibfield  {author} {\bibinfo {author} {\bibfnamefont {Y.}~\bibnamefont {Kasahara}}, \bibinfo {author} {\bibfnamefont {T.}~\bibnamefont {Ohnishi}}, \bibinfo {author} {\bibfnamefont {Y.}~\bibnamefont {Mizukami}}, \bibinfo {author} {\bibfnamefont {O.}~\bibnamefont {Tanaka}}, \bibinfo {author} {\bibfnamefont {S.}~\bibnamefont {Ma}}, \bibinfo {author} {\bibfnamefont {K.}~\bibnamefont {Sugii}}, \bibinfo {author} {\bibfnamefont {N.}~\bibnamefont {Kurita}}, \bibinfo {author} {\bibfnamefont {H.}~\bibnamefont {Tanaka}}, \bibinfo {author} {\bibfnamefont {J.}~\bibnamefont {Nasu}}, \bibinfo {author} {\bibfnamefont {Y.}~\bibnamefont {Motome}}, \bibinfo {author} {\bibfnamefont {T.}~\bibnamefont {Shibauchi}},\ and\ \bibinfo {author} {\bibfnamefont {Y.}~\bibnamefont {Matsuda}},\ }\bibfield  {title} {\bibinfo {title} {{Majorana} quantization and half-integer thermal quantum {Hall} effect in a {Kitaev} spin liquid},\ }\href {https://doi.org/10.1038/s41586-018-0274-0} {\bibfield  {journal} {\bibinfo  {journal} {Nature}\ }\textbf {\bibinfo {volume} {559}},\ \bibinfo {pages} {227} (\bibinfo {year} {2018})}\BibitemShut {NoStop}%
\bibitem [{\citenamefont {Yokoi}\ \emph {et~al.}(2021)\citenamefont {Yokoi}, \citenamefont {Ma}, \citenamefont {Kasahara}, \citenamefont {Kasahara}, \citenamefont {Shibauchi}, \citenamefont {Kurita}, \citenamefont {Tanaka}, \citenamefont {Nasu}, \citenamefont {Motome}, \citenamefont {Hickey}, \citenamefont {Trebst},\ and\ \citenamefont {Matsuda}}]{Yokoi2021}%
  \BibitemOpen
  \bibfield  {author} {\bibinfo {author} {\bibfnamefont {T.}~\bibnamefont {Yokoi}}, \bibinfo {author} {\bibfnamefont {S.}~\bibnamefont {Ma}}, \bibinfo {author} {\bibfnamefont {Y.}~\bibnamefont {Kasahara}}, \bibinfo {author} {\bibfnamefont {S.}~\bibnamefont {Kasahara}}, \bibinfo {author} {\bibfnamefont {T.}~\bibnamefont {Shibauchi}}, \bibinfo {author} {\bibfnamefont {N.}~\bibnamefont {Kurita}}, \bibinfo {author} {\bibfnamefont {H.}~\bibnamefont {Tanaka}}, \bibinfo {author} {\bibfnamefont {J.}~\bibnamefont {Nasu}}, \bibinfo {author} {\bibfnamefont {Y.}~\bibnamefont {Motome}}, \bibinfo {author} {\bibfnamefont {C.}~\bibnamefont {Hickey}}, \bibinfo {author} {\bibfnamefont {S.}~\bibnamefont {Trebst}},\ and\ \bibinfo {author} {\bibfnamefont {Y.}~\bibnamefont {Matsuda}},\ }\bibfield  {title} {\bibinfo {title} {{Half-integer quantized anomalous thermal Hall effect in the Kitaev material candidate $\alpha$-RuCl$_3$}},\ }\href {https://doi.org/https://doi.org/10.1126/science.aay5551} {\bibfield  {journal} {\bibinfo  {journal} {Science}\ }\textbf {\bibinfo {volume} {373}},\ \bibinfo {pages} {568} (\bibinfo {year} {2021})}\BibitemShut {NoStop}%
\bibitem [{\citenamefont {Tanaka}\ \emph {et~al.}(2022)\citenamefont {Tanaka}, \citenamefont {Mizukami}, \citenamefont {Harasawa}, \citenamefont {Hashimoto}, \citenamefont {Hwang}, \citenamefont {Kurita}, \citenamefont {Tanaka}, \citenamefont {Fujimoto}, \citenamefont {Matsuda}, \citenamefont {Moon},\ and\ \citenamefont {Shibauchi}}]{Tanaka2022}%
  \BibitemOpen
  \bibfield  {author} {\bibinfo {author} {\bibfnamefont {O.}~\bibnamefont {Tanaka}}, \bibinfo {author} {\bibfnamefont {Y.}~\bibnamefont {Mizukami}}, \bibinfo {author} {\bibfnamefont {R.}~\bibnamefont {Harasawa}}, \bibinfo {author} {\bibfnamefont {K.}~\bibnamefont {Hashimoto}}, \bibinfo {author} {\bibfnamefont {K.}~\bibnamefont {Hwang}}, \bibinfo {author} {\bibfnamefont {N.}~\bibnamefont {Kurita}}, \bibinfo {author} {\bibfnamefont {H.}~\bibnamefont {Tanaka}}, \bibinfo {author} {\bibfnamefont {S.}~\bibnamefont {Fujimoto}}, \bibinfo {author} {\bibfnamefont {Y.}~\bibnamefont {Matsuda}}, \bibinfo {author} {\bibfnamefont {E.-G.}\ \bibnamefont {Moon}},\ and\ \bibinfo {author} {\bibfnamefont {T.}~\bibnamefont {Shibauchi}},\ }\bibfield  {title} {\bibinfo {title} {{Thermodynamic evidence for a field-angle-dependent Majorana gap in a Kitaev spin liquid}},\ }\href {https://doi.org/https://doi.org/10.1038/s41567-021-01488-6} {\bibfield  {journal} {\bibinfo  {journal} {Nat. Phys.}\ }\textbf {\bibinfo {volume} {18}},\ \bibinfo {pages} {429} (\bibinfo {year} {2022})}\BibitemShut {NoStop}%
\bibitem [{\citenamefont {Imamura}\ \emph {et~al.}(2023)\citenamefont {Imamura}, \citenamefont {Suetsugu}, \citenamefont {Mizukami}, \citenamefont {Yoshida}, \citenamefont {Hashimoto}, \citenamefont {Ohtsuka}, \citenamefont {Kasahara}, \citenamefont {Kurita}, \citenamefont {Tanaka}, \citenamefont {Noh}, \citenamefont {Moon}, \citenamefont {Matsuda},\ and\ \citenamefont {Shibauchi}}]{Imamura2023}%
  \BibitemOpen
  \bibfield  {author} {\bibinfo {author} {\bibfnamefont {K.}~\bibnamefont {Imamura}}, \bibinfo {author} {\bibfnamefont {S.}~\bibnamefont {Suetsugu}}, \bibinfo {author} {\bibfnamefont {Y.}~\bibnamefont {Mizukami}}, \bibinfo {author} {\bibfnamefont {Y.}~\bibnamefont {Yoshida}}, \bibinfo {author} {\bibfnamefont {K.}~\bibnamefont {Hashimoto}}, \bibinfo {author} {\bibfnamefont {K.}~\bibnamefont {Ohtsuka}}, \bibinfo {author} {\bibfnamefont {Y.}~\bibnamefont {Kasahara}}, \bibinfo {author} {\bibfnamefont {N.}~\bibnamefont {Kurita}}, \bibinfo {author} {\bibfnamefont {H.}~\bibnamefont {Tanaka}}, \bibinfo {author} {\bibfnamefont {P.}~\bibnamefont {Noh}}, \bibinfo {author} {\bibfnamefont {E.-G.}\ \bibnamefont {Moon}}, \bibinfo {author} {\bibfnamefont {Y.}~\bibnamefont {Matsuda}},\ and\ \bibinfo {author} {\bibfnamefont {T.}~\bibnamefont {Shibauchi}},\ }\bibfield  {title} {\bibinfo {title} {{Majorana-fermion origin of the planar thermal Hall effect in the Kitaev magnet $\alpha$-RuCl$_3$}},\ }\href {https://doi.org/https://doi.org/10.48550/arXiv.2305.10619} {\bibfield  {journal} {\bibinfo  {journal} {preprint}\ }\textbf {\bibinfo {volume} {\!\!}},\ \bibinfo {pages} {arXiv:2305.10619} (\bibinfo {year} {2023})}\BibitemShut {NoStop}%
\bibitem [{\citenamefont {Yamashita}\ \emph {et~al.}(2020)\citenamefont {Yamashita}, \citenamefont {Gouchi}, \citenamefont {Uwatoko}, \citenamefont {Kurita},\ and\ \citenamefont {Tanaka}}]{Yamashita2020}%
  \BibitemOpen
  \bibfield  {author} {\bibinfo {author} {\bibfnamefont {M.}~\bibnamefont {Yamashita}}, \bibinfo {author} {\bibfnamefont {J.}~\bibnamefont {Gouchi}}, \bibinfo {author} {\bibfnamefont {Y.}~\bibnamefont {Uwatoko}}, \bibinfo {author} {\bibfnamefont {N.}~\bibnamefont {Kurita}},\ and\ \bibinfo {author} {\bibfnamefont {H.}~\bibnamefont {Tanaka}},\ }\bibfield  {title} {\bibinfo {title} {{Sample dependence of half-integer quantized thermal Hall effect in the Kitaev spin-liquid candidate $\alpha$-RuCl$_{3}$}},\ }\href {https://doi.org/10.1103/PhysRevB.102.220404} {\bibfield  {journal} {\bibinfo  {journal} {Phys. Rev. B}\ }\textbf {\bibinfo {volume} {102}},\ \bibinfo {pages} {220404(R)} (\bibinfo {year} {2020})}\BibitemShut {NoStop}%
\bibitem [{\citenamefont {Bruin}\ \emph {et~al.}(2022)\citenamefont {Bruin}, \citenamefont {Claus}, \citenamefont {Matsumoto}, \citenamefont {Kurita}, \citenamefont {Tanaka},\ and\ \citenamefont {Takagi}}]{Bruin2022}%
  \BibitemOpen
  \bibfield  {author} {\bibinfo {author} {\bibfnamefont {J.~A.~N.}\ \bibnamefont {Bruin}}, \bibinfo {author} {\bibfnamefont {R.~R.}\ \bibnamefont {Claus}}, \bibinfo {author} {\bibfnamefont {Y.}~\bibnamefont {Matsumoto}}, \bibinfo {author} {\bibfnamefont {N.}~\bibnamefont {Kurita}}, \bibinfo {author} {\bibfnamefont {H.}~\bibnamefont {Tanaka}},\ and\ \bibinfo {author} {\bibfnamefont {H.}~\bibnamefont {Takagi}},\ }\bibfield  {title} {\bibinfo {title} {{Robustness of the thermal Hall effect close to half-quantization in $\alpha$-RuCl$_3$}},\ }\href {https://doi.org/https://doi.org/10.1038/s41567-021-01501-y} {\bibfield  {journal} {\bibinfo  {journal} {Nat. Phys.}\ }\textbf {\bibinfo {volume} {18}},\ \bibinfo {pages} {401} (\bibinfo {year} {2022})}\BibitemShut {NoStop}%
\bibitem [{\citenamefont {Kasahara}\ \emph {et~al.}(2022)\citenamefont {Kasahara}, \citenamefont {Suetsugu}, \citenamefont {Asaba}, \citenamefont {Kasahara}, \citenamefont {Shibauchi}, \citenamefont {Kurita}, \citenamefont {Tanaka},\ and\ \citenamefont {Matsuda}}]{Kasahara2022}%
  \BibitemOpen
  \bibfield  {author} {\bibinfo {author} {\bibfnamefont {Y.}~\bibnamefont {Kasahara}}, \bibinfo {author} {\bibfnamefont {S.}~\bibnamefont {Suetsugu}}, \bibinfo {author} {\bibfnamefont {T.}~\bibnamefont {Asaba}}, \bibinfo {author} {\bibfnamefont {S.}~\bibnamefont {Kasahara}}, \bibinfo {author} {\bibfnamefont {T.}~\bibnamefont {Shibauchi}}, \bibinfo {author} {\bibfnamefont {N.}~\bibnamefont {Kurita}}, \bibinfo {author} {\bibfnamefont {H.}~\bibnamefont {Tanaka}},\ and\ \bibinfo {author} {\bibfnamefont {Y.}~\bibnamefont {Matsuda}},\ }\bibfield  {title} {\bibinfo {title} {{Quantized and unquantized thermal Hall conductance of the Kitaev spin liquid candidate $\alpha$-RuCl$_3$}},\ }\href {https://doi.org/10.1103/PhysRevB.106.L060410} {\bibfield  {journal} {\bibinfo  {journal} {Phys. Rev. B}\ }\textbf {\bibinfo {volume} {106}},\ \bibinfo {pages} {L060410} (\bibinfo {year} {2022})}\BibitemShut {NoStop}%
\bibitem [{\citenamefont {Czajka}\ \emph {et~al.}(2023)\citenamefont {Czajka}, \citenamefont {Gao}, \citenamefont {Hirschberger}, \citenamefont {Lampen-Kelley}, \citenamefont {Banerjee}, \citenamefont {Quirk}, \citenamefont {Mandrus}, \citenamefont {Nagler},\ and\ \citenamefont {Ong}}]{Czajka2023}%
  \BibitemOpen
  \bibfield  {author} {\bibinfo {author} {\bibfnamefont {P.}~\bibnamefont {Czajka}}, \bibinfo {author} {\bibfnamefont {T.}~\bibnamefont {Gao}}, \bibinfo {author} {\bibfnamefont {M.}~\bibnamefont {Hirschberger}}, \bibinfo {author} {\bibfnamefont {P.}~\bibnamefont {Lampen-Kelley}}, \bibinfo {author} {\bibfnamefont {A.}~\bibnamefont {Banerjee}}, \bibinfo {author} {\bibfnamefont {N.}~\bibnamefont {Quirk}}, \bibinfo {author} {\bibfnamefont {D.~G.}\ \bibnamefont {Mandrus}}, \bibinfo {author} {\bibfnamefont {S.~E.}\ \bibnamefont {Nagler}},\ and\ \bibinfo {author} {\bibfnamefont {N.~P.}\ \bibnamefont {Ong}},\ }\bibfield  {title} {\bibinfo {title} {{Planar thermal Hall effect of topological bosons in the Kitaev magnet $\alpha$-RuCl$_3$}},\ }\href {https://doi.org/10.1038/s41563-022-01397-w} {\bibfield  {journal} {\bibinfo  {journal} {Nat. Mater.}\ }\textbf {\bibinfo {volume} {22}},\ \bibinfo {pages} {36} (\bibinfo {year} {2023})}\BibitemShut {NoStop}%
\bibitem [{\citenamefont {McClarty}\ \emph {et~al.}(2018)\citenamefont {McClarty}, \citenamefont {Dong}, \citenamefont {Gohlke}, \citenamefont {Rau}, \citenamefont {Pollmann}, \citenamefont {Moessner},\ and\ \citenamefont {Penc}}]{McClarty2018}%
  \BibitemOpen
  \bibfield  {author} {\bibinfo {author} {\bibfnamefont {P.~A.}\ \bibnamefont {McClarty}}, \bibinfo {author} {\bibfnamefont {X.-Y.}\ \bibnamefont {Dong}}, \bibinfo {author} {\bibfnamefont {M.}~\bibnamefont {Gohlke}}, \bibinfo {author} {\bibfnamefont {J.~G.}\ \bibnamefont {Rau}}, \bibinfo {author} {\bibfnamefont {F.}~\bibnamefont {Pollmann}}, \bibinfo {author} {\bibfnamefont {R.}~\bibnamefont {Moessner}},\ and\ \bibinfo {author} {\bibfnamefont {K.}~\bibnamefont {Penc}},\ }\bibfield  {title} {\bibinfo {title} {Topological magnons in kitaev magnets at high fields},\ }\href {https://doi.org/10.1103/PhysRevB.98.060404} {\bibfield  {journal} {\bibinfo  {journal} {Phys. Rev. B}\ }\textbf {\bibinfo {volume} {98}},\ \bibinfo {pages} {060404} (\bibinfo {year} {2018})}\BibitemShut {NoStop}%
\bibitem [{\citenamefont {Joshi}(2018)}]{Joshi2018}%
  \BibitemOpen
  \bibfield  {author} {\bibinfo {author} {\bibfnamefont {D.~G.}\ \bibnamefont {Joshi}},\ }\bibfield  {title} {\bibinfo {title} {Topological excitations in the ferromagnetic kitaev-heisenberg model},\ }\href {https://doi.org/10.1103/PhysRevB.98.060405} {\bibfield  {journal} {\bibinfo  {journal} {Phys. Rev. B}\ }\textbf {\bibinfo {volume} {98}},\ \bibinfo {pages} {060405} (\bibinfo {year} {2018})}\BibitemShut {NoStop}%
\bibitem [{\citenamefont {Chern}\ \emph {et~al.}(2021)\citenamefont {Chern}, \citenamefont {Zhang},\ and\ \citenamefont {Kim}}]{Chern2021}%
  \BibitemOpen
  \bibfield  {author} {\bibinfo {author} {\bibfnamefont {L.~E.}\ \bibnamefont {Chern}}, \bibinfo {author} {\bibfnamefont {E.~Z.}\ \bibnamefont {Zhang}},\ and\ \bibinfo {author} {\bibfnamefont {Y.~B.}\ \bibnamefont {Kim}},\ }\bibfield  {title} {\bibinfo {title} {Sign structure of thermal hall conductivity and topological magnons for in-plane field polarized kitaev magnets},\ }\href {https://doi.org/10.1103/PhysRevLett.126.147201} {\bibfield  {journal} {\bibinfo  {journal} {Phys. Rev. Lett.}\ }\textbf {\bibinfo {volume} {126}},\ \bibinfo {pages} {147201} (\bibinfo {year} {2021})}\BibitemShut {NoStop}%
\bibitem [{\citenamefont {Zhang}\ \emph {et~al.}(2021)\citenamefont {Zhang}, \citenamefont {Chern},\ and\ \citenamefont {Kim}}]{Zhang2021}%
  \BibitemOpen
  \bibfield  {author} {\bibinfo {author} {\bibfnamefont {E.~Z.}\ \bibnamefont {Zhang}}, \bibinfo {author} {\bibfnamefont {L.~E.}\ \bibnamefont {Chern}},\ and\ \bibinfo {author} {\bibfnamefont {Y.~B.}\ \bibnamefont {Kim}},\ }\bibfield  {title} {\bibinfo {title} {Topological magnons for thermal hall transport in frustrated magnets with bond-dependent interactions},\ }\href {https://doi.org/10.1103/PhysRevB.103.174402} {\bibfield  {journal} {\bibinfo  {journal} {Phys. Rev. B}\ }\textbf {\bibinfo {volume} {103}},\ \bibinfo {pages} {174402} (\bibinfo {year} {2021})}\BibitemShut {NoStop}%
\bibitem [{\citenamefont {Lefran\ifmmode~\mbox{\c{c}}\else \c{c}\fi{}ois}\ \emph {et~al.}(2022)\citenamefont {Lefran\ifmmode~\mbox{\c{c}}\else \c{c}\fi{}ois}, \citenamefont {Grissonnanche}, \citenamefont {Baglo}, \citenamefont {Lampen-Kelley}, \citenamefont {Yan}, \citenamefont {Balz}, \citenamefont {Mandrus}, \citenamefont {Nagler}, \citenamefont {Kim}, \citenamefont {Kim}, \citenamefont {Doiron-Leyraud},\ and\ \citenamefont {Taillefer}}]{lefranccois2022}%
  \BibitemOpen
  \bibfield  {author} {\bibinfo {author} {\bibfnamefont {E.}~\bibnamefont {Lefran\ifmmode~\mbox{\c{c}}\else \c{c}\fi{}ois}}, \bibinfo {author} {\bibfnamefont {G.}~\bibnamefont {Grissonnanche}}, \bibinfo {author} {\bibfnamefont {J.}~\bibnamefont {Baglo}}, \bibinfo {author} {\bibfnamefont {P.}~\bibnamefont {Lampen-Kelley}}, \bibinfo {author} {\bibfnamefont {J.-Q.}\ \bibnamefont {Yan}}, \bibinfo {author} {\bibfnamefont {C.}~\bibnamefont {Balz}}, \bibinfo {author} {\bibfnamefont {D.}~\bibnamefont {Mandrus}}, \bibinfo {author} {\bibfnamefont {S.~E.}\ \bibnamefont {Nagler}}, \bibinfo {author} {\bibfnamefont {S.}~\bibnamefont {Kim}}, \bibinfo {author} {\bibfnamefont {Y.-J.}\ \bibnamefont {Kim}}, \bibinfo {author} {\bibfnamefont {N.}~\bibnamefont {Doiron-Leyraud}},\ and\ \bibinfo {author} {\bibfnamefont {L.}~\bibnamefont {Taillefer}},\ }\bibfield  {title} {\bibinfo {title} {Evidence of a phonon hall effect in the kitaev spin liquid candidate $\ensuremath{\alpha}\text{\ensuremath{-}}{\mathrm{rucl}}_{3}$},\ }\href {https://doi.org/10.1103/PhysRevX.12.021025} {\bibfield  {journal} {\bibinfo  {journal} {Phys. Rev. X}\ }\textbf {\bibinfo {volume} {12}},\ \bibinfo {pages} {021025} (\bibinfo {year} {2022})}\BibitemShut {NoStop}%
\bibitem [{\citenamefont {Lampen-Kelley}\ \emph {et~al.}(2017)\citenamefont {Lampen-Kelley}, \citenamefont {Banerjee}, \citenamefont {Aczel}, \citenamefont {Cao}, \citenamefont {Stone}, \citenamefont {Bridges}, \citenamefont {Yan}, \citenamefont {Nagler},\ and\ \citenamefont {Mandrus}}]{Lampen-Kelley2017}%
  \BibitemOpen
  \bibfield  {author} {\bibinfo {author} {\bibfnamefont {P.}~\bibnamefont {Lampen-Kelley}}, \bibinfo {author} {\bibfnamefont {A.}~\bibnamefont {Banerjee}}, \bibinfo {author} {\bibfnamefont {A.~A.}\ \bibnamefont {Aczel}}, \bibinfo {author} {\bibfnamefont {H.~B.}\ \bibnamefont {Cao}}, \bibinfo {author} {\bibfnamefont {M.~B.}\ \bibnamefont {Stone}}, \bibinfo {author} {\bibfnamefont {C.~A.}\ \bibnamefont {Bridges}}, \bibinfo {author} {\bibfnamefont {J.-Q.}\ \bibnamefont {Yan}}, \bibinfo {author} {\bibfnamefont {S.~E.}\ \bibnamefont {Nagler}},\ and\ \bibinfo {author} {\bibfnamefont {D.}~\bibnamefont {Mandrus}},\ }\bibfield  {title} {\bibinfo {title} {{Destabilization of Magnetic Order in a Dilute Kitaev Spin Liquid Candidate}},\ }\href {https://doi.org/10.1103/PhysRevLett.119.237203} {\bibfield  {journal} {\bibinfo  {journal} {Phys. Rev. Lett.}\ }\textbf {\bibinfo {volume} {119}},\ \bibinfo {pages} {237203} (\bibinfo {year} {2017})}\BibitemShut {NoStop}%
\bibitem [{\citenamefont {Do}\ \emph {et~al.}(2018)\citenamefont {Do}, \citenamefont {Lee}, \citenamefont {Lee}, \citenamefont {Choi}, \citenamefont {Lee}, \citenamefont {Gorbunov}, \citenamefont {Wosnitza}, \citenamefont {Suh},\ and\ \citenamefont {Choi}}]{Do2018}%
  \BibitemOpen
  \bibfield  {author} {\bibinfo {author} {\bibfnamefont {S.-H.}\ \bibnamefont {Do}}, \bibinfo {author} {\bibfnamefont {W.-J.}\ \bibnamefont {Lee}}, \bibinfo {author} {\bibfnamefont {S.}~\bibnamefont {Lee}}, \bibinfo {author} {\bibfnamefont {Y.~S.}\ \bibnamefont {Choi}}, \bibinfo {author} {\bibfnamefont {K.-J.}\ \bibnamefont {Lee}}, \bibinfo {author} {\bibfnamefont {D.~I.}\ \bibnamefont {Gorbunov}}, \bibinfo {author} {\bibfnamefont {J.}~\bibnamefont {Wosnitza}}, \bibinfo {author} {\bibfnamefont {B.~J.}\ \bibnamefont {Suh}},\ and\ \bibinfo {author} {\bibfnamefont {K.-Y.}\ \bibnamefont {Choi}},\ }\bibfield  {title} {\bibinfo {title} {{Short-range quasistatic order and critical spin correlations in $\alpha$-Ru$_{1-x}$Ir$_x$Cl$_{3}$}},\ }\href {https://doi.org/10.1103/PhysRevB.98.014407} {\bibfield  {journal} {\bibinfo  {journal} {Phys. Rev. B}\ }\textbf {\bibinfo {volume} {98}},\ \bibinfo {pages} {014407} (\bibinfo {year} {2018})}\BibitemShut {NoStop}%
\bibitem [{\citenamefont {Do}\ \emph {et~al.}(2020)\citenamefont {Do}, \citenamefont {Lee}, \citenamefont {Kihara}, \citenamefont {Choi}, \citenamefont {Yoon}, \citenamefont {Kim}, \citenamefont {Cheong}, \citenamefont {Chen}, \citenamefont {Chou}, \citenamefont {Nojiri},\ and\ \citenamefont {Choi}}]{Do2020}%
  \BibitemOpen
  \bibfield  {author} {\bibinfo {author} {\bibfnamefont {S.-H.}\ \bibnamefont {Do}}, \bibinfo {author} {\bibfnamefont {C.~H.}\ \bibnamefont {Lee}}, \bibinfo {author} {\bibfnamefont {T.}~\bibnamefont {Kihara}}, \bibinfo {author} {\bibfnamefont {Y.~S.}\ \bibnamefont {Choi}}, \bibinfo {author} {\bibfnamefont {S.}~\bibnamefont {Yoon}}, \bibinfo {author} {\bibfnamefont {K.}~\bibnamefont {Kim}}, \bibinfo {author} {\bibfnamefont {H.}~\bibnamefont {Cheong}}, \bibinfo {author} {\bibfnamefont {W.-T.}\ \bibnamefont {Chen}}, \bibinfo {author} {\bibfnamefont {F.}~\bibnamefont {Chou}}, \bibinfo {author} {\bibfnamefont {H.}~\bibnamefont {Nojiri}},\ and\ \bibinfo {author} {\bibfnamefont {K.-Y.}\ \bibnamefont {Choi}},\ }\bibfield  {title} {\bibinfo {title} {{Randomly Hopping Majorana Fermions in the Diluted Kitaev System $\alpha$-Ru$_{0.8}$Ir$_{0.2}$Cl$_{3}$}},\ }\href {https://doi.org/10.1103/PhysRevLett.124.047204} {\bibfield  {journal} {\bibinfo  {journal} {Phys. Rev. Lett.}\ }\textbf {\bibinfo {volume} {124}},\ \bibinfo {pages} {047204} (\bibinfo {year} {2020})}\BibitemShut {NoStop}%
\bibitem [{\citenamefont {Kitagawa}\ \emph {et~al.}(2018)\citenamefont {Kitagawa}, \citenamefont {Takayama}, \citenamefont {Matsumoto}, \citenamefont {Kato}, \citenamefont {Takano}, \citenamefont {Kishimoto}, \citenamefont {Bette}, \citenamefont {Dinnebier}, \citenamefont {Jackeli},\ and\ \citenamefont {Takagi}}]{Kitagawa2018}%
  \BibitemOpen
  \bibfield  {author} {\bibinfo {author} {\bibfnamefont {K.}~\bibnamefont {Kitagawa}}, \bibinfo {author} {\bibfnamefont {T.}~\bibnamefont {Takayama}}, \bibinfo {author} {\bibfnamefont {Y.}~\bibnamefont {Matsumoto}}, \bibinfo {author} {\bibfnamefont {A.}~\bibnamefont {Kato}}, \bibinfo {author} {\bibfnamefont {R.}~\bibnamefont {Takano}}, \bibinfo {author} {\bibfnamefont {Y.}~\bibnamefont {Kishimoto}}, \bibinfo {author} {\bibfnamefont {S.}~\bibnamefont {Bette}}, \bibinfo {author} {\bibfnamefont {R.}~\bibnamefont {Dinnebier}}, \bibinfo {author} {\bibfnamefont {G.}~\bibnamefont {Jackeli}},\ and\ \bibinfo {author} {\bibfnamefont {H.}~\bibnamefont {Takagi}},\ }\bibfield  {title} {\bibinfo {title} {A spin–orbital-entangled quantum liquid on a honeycomb lattice},\ }\href {https://doi.org/10.1038/nature25482} {\bibfield  {journal} {\bibinfo  {journal} {Nature}\ }\textbf {\bibinfo {volume} {554}},\ \bibinfo {pages} {341} (\bibinfo {year} {2018})}\BibitemShut {NoStop}%
\bibitem [{\citenamefont {Willans}\ \emph {et~al.}(2010)\citenamefont {Willans}, \citenamefont {Chalker},\ and\ \citenamefont {Moessner}}]{Willans2010}%
  \BibitemOpen
  \bibfield  {author} {\bibinfo {author} {\bibfnamefont {A.~J.}\ \bibnamefont {Willans}}, \bibinfo {author} {\bibfnamefont {J.~T.}\ \bibnamefont {Chalker}},\ and\ \bibinfo {author} {\bibfnamefont {R.}~\bibnamefont {Moessner}},\ }\bibfield  {title} {\bibinfo {title} {{Disorder in a Quantum Spin Liquid: Flux Binding and Local Moment Formation}},\ }\href {https://doi.org/10.1103/PhysRevLett.104.237203} {\bibfield  {journal} {\bibinfo  {journal} {Phys. Rev. Lett.}\ }\textbf {\bibinfo {volume} {104}},\ \bibinfo {pages} {237203} (\bibinfo {year} {2010})}\BibitemShut {NoStop}%
\bibitem [{\citenamefont {Willans}\ \emph {et~al.}(2011)\citenamefont {Willans}, \citenamefont {Chalker},\ and\ \citenamefont {Moessner}}]{Willans2011}%
  \BibitemOpen
  \bibfield  {author} {\bibinfo {author} {\bibfnamefont {A.~J.}\ \bibnamefont {Willans}}, \bibinfo {author} {\bibfnamefont {J.~T.}\ \bibnamefont {Chalker}},\ and\ \bibinfo {author} {\bibfnamefont {R.}~\bibnamefont {Moessner}},\ }\bibfield  {title} {\bibinfo {title} {{Site dilution in the Kitaev honeycomb model}},\ }\href {https://doi.org/10.1103/PhysRevB.84.115146} {\bibfield  {journal} {\bibinfo  {journal} {Phys. Rev. B}\ }\textbf {\bibinfo {volume} {84}},\ \bibinfo {pages} {115146} (\bibinfo {year} {2011})}\BibitemShut {NoStop}%
\bibitem [{\citenamefont {Knolle}\ \emph {et~al.}(2019)\citenamefont {Knolle}, \citenamefont {Moessner},\ and\ \citenamefont {Perkins}}]{Knolle2019}%
  \BibitemOpen
  \bibfield  {author} {\bibinfo {author} {\bibfnamefont {J.}~\bibnamefont {Knolle}}, \bibinfo {author} {\bibfnamefont {R.}~\bibnamefont {Moessner}},\ and\ \bibinfo {author} {\bibfnamefont {N.~B.}\ \bibnamefont {Perkins}},\ }\bibfield  {title} {\bibinfo {title} {{Bond-Disordered Spin Liquid and the Honeycomb Iridate H$_{3}$LiIr$_{2}$O$_{6}$: Abundant Low-Energy Density of States from Random Majorana Hopping}},\ }\href {https://doi.org/10.1103/PhysRevLett.122.047202} {\bibfield  {journal} {\bibinfo  {journal} {Phys. Rev. Lett.}\ }\textbf {\bibinfo {volume} {122}},\ \bibinfo {pages} {047202} (\bibinfo {year} {2019})}\BibitemShut {NoStop}%
\bibitem [{\citenamefont {Nasu}\ and\ \citenamefont {Motome}(2020)}]{Nasu2020}%
  \BibitemOpen
  \bibfield  {author} {\bibinfo {author} {\bibfnamefont {J.}~\bibnamefont {Nasu}}\ and\ \bibinfo {author} {\bibfnamefont {Y.}~\bibnamefont {Motome}},\ }\bibfield  {title} {\bibinfo {title} {{Thermodynamic and transport properties in disordered Kitaev models}},\ }\href {https://doi.org/10.1103/PhysRevB.102.054437} {\bibfield  {journal} {\bibinfo  {journal} {Phys. Rev. B}\ }\textbf {\bibinfo {volume} {102}},\ \bibinfo {pages} {054437} (\bibinfo {year} {2020})}\BibitemShut {NoStop}%
\bibitem [{\citenamefont {G.}\ \emph {et~al.}(2012)\citenamefont {G.}, \citenamefont {Sreenath}, \citenamefont {Lakshminarayan},\ and\ \citenamefont {Narayanan}}]{Sreenath2012}%
  \BibitemOpen
  \bibfield  {author} {\bibinfo {author} {\bibfnamefont {S.}~\bibnamefont {G.}}, \bibinfo {author} {\bibfnamefont {V.}~\bibnamefont {Sreenath}}, \bibinfo {author} {\bibfnamefont {A.}~\bibnamefont {Lakshminarayan}},\ and\ \bibinfo {author} {\bibfnamefont {R.}~\bibnamefont {Narayanan}},\ }\bibfield  {title} {\bibinfo {title} {{Localized zero-energy modes in the Kitaev model with vacancy disorder}},\ }\href {https://doi.org/10.1103/PhysRevB.85.054204} {\bibfield  {journal} {\bibinfo  {journal} {Phys. Rev. B}\ }\textbf {\bibinfo {volume} {85}},\ \bibinfo {pages} {054204} (\bibinfo {year} {2012})}\BibitemShut {NoStop}%
\bibitem [{\citenamefont {Petrova}\ \emph {et~al.}(2013)\citenamefont {Petrova}, \citenamefont {Mellado},\ and\ \citenamefont {Tchernyshyov}}]{Petrova2013}%
  \BibitemOpen
  \bibfield  {author} {\bibinfo {author} {\bibfnamefont {O.}~\bibnamefont {Petrova}}, \bibinfo {author} {\bibfnamefont {P.}~\bibnamefont {Mellado}},\ and\ \bibinfo {author} {\bibfnamefont {O.}~\bibnamefont {Tchernyshyov}},\ }\bibfield  {title} {\bibinfo {title} {{Unpaired Majorana modes in the gapped phase of Kitaev's honeycomb model}},\ }\href {https://doi.org/10.1103/PhysRevB.88.140405} {\bibfield  {journal} {\bibinfo  {journal} {Phys. Rev. B}\ }\textbf {\bibinfo {volume} {88}},\ \bibinfo {pages} {140405(R)} (\bibinfo {year} {2013})}\BibitemShut {NoStop}%
\bibitem [{\citenamefont {Udagawa}(2018)}]{Udagawa2018}%
  \BibitemOpen
  \bibfield  {author} {\bibinfo {author} {\bibfnamefont {M.}~\bibnamefont {Udagawa}},\ }\bibfield  {title} {\bibinfo {title} {{Vison-Majorana complex zero-energy resonance in the Kitaev spin liquid}},\ }\href {https://doi.org/10.1103/PhysRevB.98.220404} {\bibfield  {journal} {\bibinfo  {journal} {Phys. Rev. B}\ }\textbf {\bibinfo {volume} {98}},\ \bibinfo {pages} {220404(R)} (\bibinfo {year} {2018})}\BibitemShut {NoStop}%
\bibitem [{\citenamefont {Dantas}\ and\ \citenamefont {Andrade}(2022)}]{Dantas2022}%
  \BibitemOpen
  \bibfield  {author} {\bibinfo {author} {\bibfnamefont {V.}~\bibnamefont {Dantas}}\ and\ \bibinfo {author} {\bibfnamefont {E.~C.}\ \bibnamefont {Andrade}},\ }\bibfield  {title} {\bibinfo {title} {Disorder, low-energy excitations, and topology in the {Kitaev} spin liquid},\ }\href {https://doi.org/10.1103/PhysRevLett.129.037204} {\bibfield  {journal} {\bibinfo  {journal} {Phys. Rev. Lett.}\ }\textbf {\bibinfo {volume} {129}},\ \bibinfo {pages} {037204} (\bibinfo {year} {2022})}\BibitemShut {NoStop}%
\bibitem [{\citenamefont {Singhania}\ \emph {et~al.}(2023)\citenamefont {Singhania}, \citenamefont {van~den Brink},\ and\ \citenamefont {Nishimoto}}]{Singhania2023}%
  \BibitemOpen
  \bibfield  {author} {\bibinfo {author} {\bibfnamefont {A.}~\bibnamefont {Singhania}}, \bibinfo {author} {\bibfnamefont {J.}~\bibnamefont {van~den Brink}},\ and\ \bibinfo {author} {\bibfnamefont {S.}~\bibnamefont {Nishimoto}},\ }\bibfield  {title} {\bibinfo {title} {{Disorder effects in the Kitaev-Heisenberg model}},\ }\href {https://doi.org/10.1103/PhysRevResearch.5.023009} {\bibfield  {journal} {\bibinfo  {journal} {Phys. Rev. Res.}\ }\textbf {\bibinfo {volume} {5}},\ \bibinfo {pages} {023009} (\bibinfo {year} {2023})}\BibitemShut {NoStop}%
\bibitem [{\citenamefont {Takahashi}\ \emph {et~al.}(2022)\citenamefont {Takahashi}, \citenamefont {Yamada}, \citenamefont {Udagawa},\ and\ \citenamefont {Fujimoto}}]{Takahashi2022}%
  \BibitemOpen
  \bibfield  {author} {\bibinfo {author} {\bibfnamefont {M.~O.}\ \bibnamefont {Takahashi}}, \bibinfo {author} {\bibfnamefont {M.~G.}\ \bibnamefont {Yamada}}, \bibinfo {author} {\bibfnamefont {M.}~\bibnamefont {Udagawa}},\ and\ \bibinfo {author} {\bibfnamefont {S.}~\bibnamefont {Fujimoto}},\ }\bibfield  {title} {\bibinfo {title} {{Non-local spin correlation as a signature of Ising anyons trapped in vacancies of the Kitaev spin liquid}},\ }\href {https://doi.org/https://doi.org/10.48550/arXiv.2211.13884} {\bibfield  {journal} {\bibinfo  {journal} {preprint}\ }\textbf {\bibinfo {volume} {\!\!}},\ \bibinfo {pages} {arXiv:2211.13884} (\bibinfo {year} {2022})}\BibitemShut {NoStop}%
\bibitem [{\citenamefont {Yamada}(2020)}]{Yamada2020}%
  \BibitemOpen
  \bibfield  {author} {\bibinfo {author} {\bibfnamefont {M.~G.}\ \bibnamefont {Yamada}},\ }\bibfield  {title} {\bibinfo {title} {{Anderson-Kitaev spin liquid}},\ }\href {https://doi.org/https://doi.org/10.1038/s41535-020-00285-3} {\bibfield  {journal} {\bibinfo  {journal} {npj Quantum Mater.}\ }\textbf {\bibinfo {volume} {5}},\ \bibinfo {pages} {82} (\bibinfo {year} {2020})}\BibitemShut {NoStop}%
\bibitem [{\citenamefont {Kao}\ \emph {et~al.}(2021)\citenamefont {Kao}, \citenamefont {Knolle}, \citenamefont {Hal\'asz}, \citenamefont {Moessner},\ and\ \citenamefont {Perkins}}]{Kao2021}%
  \BibitemOpen
  \bibfield  {author} {\bibinfo {author} {\bibfnamefont {W.-H.}\ \bibnamefont {Kao}}, \bibinfo {author} {\bibfnamefont {J.}~\bibnamefont {Knolle}}, \bibinfo {author} {\bibfnamefont {G.~B.}\ \bibnamefont {Hal\'asz}}, \bibinfo {author} {\bibfnamefont {R.}~\bibnamefont {Moessner}},\ and\ \bibinfo {author} {\bibfnamefont {N.~B.}\ \bibnamefont {Perkins}},\ }\bibfield  {title} {\bibinfo {title} {{Vacancy-Induced Low-Energy Density of States in the Kitaev Spin Liquid}},\ }\href {https://doi.org/10.1103/PhysRevX.11.011034} {\bibfield  {journal} {\bibinfo  {journal} {Phys. Rev. X}\ }\textbf {\bibinfo {volume} {11}},\ \bibinfo {pages} {011034} (\bibinfo {year} {2021})}\BibitemShut {NoStop}%
\bibitem [{\citenamefont {Mizukami}\ \emph {et~al.}(2014)\citenamefont {Mizukami}, \citenamefont {Konczykowski}, \citenamefont {Kawamoto}, \citenamefont {Kurata}, \citenamefont {Kasahara}, \citenamefont {Hashimoto}, \citenamefont {Mishra}, \citenamefont {Kreisel}, \citenamefont {Wang}, \citenamefont {Hirschfeld} \emph {et~al.}}]{Mizukami2014}%
  \BibitemOpen
  \bibfield  {author} {\bibinfo {author} {\bibfnamefont {Y.}~\bibnamefont {Mizukami}}, \bibinfo {author} {\bibfnamefont {M.}~\bibnamefont {Konczykowski}}, \bibinfo {author} {\bibfnamefont {Y.}~\bibnamefont {Kawamoto}}, \bibinfo {author} {\bibfnamefont {S.}~\bibnamefont {Kurata}}, \bibinfo {author} {\bibfnamefont {S.}~\bibnamefont {Kasahara}}, \bibinfo {author} {\bibfnamefont {K.}~\bibnamefont {Hashimoto}}, \bibinfo {author} {\bibfnamefont {V.}~\bibnamefont {Mishra}}, \bibinfo {author} {\bibfnamefont {A.}~\bibnamefont {Kreisel}}, \bibinfo {author} {\bibfnamefont {Y.}~\bibnamefont {Wang}}, \bibinfo {author} {\bibfnamefont {P.}~\bibnamefont {Hirschfeld}}, \emph {et~al.},\ }\bibfield  {title} {\bibinfo {title} {{Disorder-induced topological change of the superconducting gap structure in iron pnictides}},\ }\href {https://doi.org/https://doi.org/10.1038/ncomms6657} {\bibfield  {journal} {\bibinfo  {journal} {Nat. Commun.}\ }\textbf {\bibinfo {volume} {5}},\ \bibinfo {pages} {5657} (\bibinfo {year} {2014})}\BibitemShut {NoStop}%
\bibitem [{\citenamefont {Cho}\ \emph {et~al.}(2018)\citenamefont {Cho}, \citenamefont {Ko{\'n}czykowski}, \citenamefont {Teknowijoyo}, \citenamefont {Tanatar},\ and\ \citenamefont {Prozorov}}]{Cho2018}%
  \BibitemOpen
  \bibfield  {author} {\bibinfo {author} {\bibfnamefont {K.}~\bibnamefont {Cho}}, \bibinfo {author} {\bibfnamefont {M.}~\bibnamefont {Ko{\'n}czykowski}}, \bibinfo {author} {\bibfnamefont {S.}~\bibnamefont {Teknowijoyo}}, \bibinfo {author} {\bibfnamefont {M.~A.}\ \bibnamefont {Tanatar}},\ and\ \bibinfo {author} {\bibfnamefont {R.}~\bibnamefont {Prozorov}},\ }\bibfield  {title} {\bibinfo {title} {{Using electron irradiation to probe iron-based superconductors}},\ }\href {https://doi.org/https://doi.org/10.1088/1361-6668/aabfa8} {\bibfield  {journal} {\bibinfo  {journal} {Supercond. Sci. Technol.}\ }\textbf {\bibinfo {volume} {31}},\ \bibinfo {pages} {064002} (\bibinfo {year} {2018})}\BibitemShut {NoStop}%
\bibitem [{\citenamefont {Kubota}\ \emph {et~al.}(2015)\citenamefont {Kubota}, \citenamefont {Tanaka}, \citenamefont {Ono}, \citenamefont {Narumi},\ and\ \citenamefont {Kindo}}]{Kubota2015}%
  \BibitemOpen
  \bibfield  {author} {\bibinfo {author} {\bibfnamefont {Y.}~\bibnamefont {Kubota}}, \bibinfo {author} {\bibfnamefont {H.}~\bibnamefont {Tanaka}}, \bibinfo {author} {\bibfnamefont {T.}~\bibnamefont {Ono}}, \bibinfo {author} {\bibfnamefont {Y.}~\bibnamefont {Narumi}},\ and\ \bibinfo {author} {\bibfnamefont {K.}~\bibnamefont {Kindo}},\ }\bibfield  {title} {\bibinfo {title} {{Successive magnetic phase transitions in $\alpha$-{RuCl}$_{3}$: XY-like frustrated magnet on the honeycomb lattice}},\ }\href {https://doi.org/10.1103/PhysRevB.91.094422} {\bibfield  {journal} {\bibinfo  {journal} {Phys. Rev. B}\ }\textbf {\bibinfo {volume} {91}},\ \bibinfo {pages} {094422} (\bibinfo {year} {2015})}\BibitemShut {NoStop}%
\bibitem [{\citenamefont {Cao}\ \emph {et~al.}(2016)\citenamefont {Cao}, \citenamefont {Banerjee}, \citenamefont {Yan}, \citenamefont {Bridges}, \citenamefont {Lumsden}, \citenamefont {Mandrus}, \citenamefont {Tennant}, \citenamefont {Chakoumakos},\ and\ \citenamefont {Nagler}}]{Cao2016}%
  \BibitemOpen
  \bibfield  {author} {\bibinfo {author} {\bibfnamefont {H.~B.}\ \bibnamefont {Cao}}, \bibinfo {author} {\bibfnamefont {A.}~\bibnamefont {Banerjee}}, \bibinfo {author} {\bibfnamefont {J.-Q.}\ \bibnamefont {Yan}}, \bibinfo {author} {\bibfnamefont {C.~A.}\ \bibnamefont {Bridges}}, \bibinfo {author} {\bibfnamefont {M.~D.}\ \bibnamefont {Lumsden}}, \bibinfo {author} {\bibfnamefont {D.~G.}\ \bibnamefont {Mandrus}}, \bibinfo {author} {\bibfnamefont {D.~A.}\ \bibnamefont {Tennant}}, \bibinfo {author} {\bibfnamefont {B.~C.}\ \bibnamefont {Chakoumakos}},\ and\ \bibinfo {author} {\bibfnamefont {S.~E.}\ \bibnamefont {Nagler}},\ }\bibfield  {title} {\bibinfo {title} {{Low-temperature crystal and magnetic structure of $\alpha$-{RuCl}$_{3}$}},\ }\href {https://doi.org/10.1103/PhysRevB.93.134423} {\bibfield  {journal} {\bibinfo  {journal} {Phys. Rev. B}\ }\textbf {\bibinfo {volume} {93}},\ \bibinfo {pages} {134423} (\bibinfo {year} {2016})}\BibitemShut {NoStop}%
\bibitem [{\citenamefont {Widmann}\ \emph {et~al.}(2019)\citenamefont {Widmann}, \citenamefont {Tsurkan}, \citenamefont {Prishchenko}, \citenamefont {Mazurenko}, \citenamefont {Tsirlin},\ and\ \citenamefont {Loidl}}]{Widmann2019}%
  \BibitemOpen
  \bibfield  {author} {\bibinfo {author} {\bibfnamefont {S.}~\bibnamefont {Widmann}}, \bibinfo {author} {\bibfnamefont {V.}~\bibnamefont {Tsurkan}}, \bibinfo {author} {\bibfnamefont {D.~A.}\ \bibnamefont {Prishchenko}}, \bibinfo {author} {\bibfnamefont {V.~G.}\ \bibnamefont {Mazurenko}}, \bibinfo {author} {\bibfnamefont {A.~A.}\ \bibnamefont {Tsirlin}},\ and\ \bibinfo {author} {\bibfnamefont {A.}~\bibnamefont {Loidl}},\ }\bibfield  {title} {\bibinfo {title} {Thermodynamic evidence of fractionalized excitations in $\alpha$-{RuCl}$_3$},\ }\href {https://doi.org/10.1103/PhysRevB.99.094415} {\bibfield  {journal} {\bibinfo  {journal} {Phys. Rev. B}\ }\textbf {\bibinfo {volume} {99}},\ \bibinfo {pages} {094415} (\bibinfo {year} {2019})}\BibitemShut {NoStop}%
\bibitem [{\citenamefont {Sears}\ \emph {et~al.}(2017)\citenamefont {Sears}, \citenamefont {Zhao}, \citenamefont {Xu}, \citenamefont {Lynn},\ and\ \citenamefont {Kim}}]{Sears2017}%
  \BibitemOpen
  \bibfield  {author} {\bibinfo {author} {\bibfnamefont {J.~A.}\ \bibnamefont {Sears}}, \bibinfo {author} {\bibfnamefont {Y.}~\bibnamefont {Zhao}}, \bibinfo {author} {\bibfnamefont {Z.}~\bibnamefont {Xu}}, \bibinfo {author} {\bibfnamefont {J.~W.}\ \bibnamefont {Lynn}},\ and\ \bibinfo {author} {\bibfnamefont {Y.-J.}\ \bibnamefont {Kim}},\ }\bibfield  {title} {\bibinfo {title} {{Phase diagram of $\alpha$-{RuCl}$_{3}$ in an in-plane magnetic field}},\ }\href {https://doi.org/10.1103/PhysRevB.95.180411} {\bibfield  {journal} {\bibinfo  {journal} {Phys. Rev. B}\ }\textbf {\bibinfo {volume} {95}},\ \bibinfo {pages} {180411(R)} (\bibinfo {year} {2017})}\BibitemShut {NoStop}%
\bibitem [{\citenamefont {Wolter}\ \emph {et~al.}(2017)\citenamefont {Wolter}, \citenamefont {Corredor}, \citenamefont {Janssen}, \citenamefont {Nenkov}, \citenamefont {Sch\"onecker}, \citenamefont {Do}, \citenamefont {Choi}, \citenamefont {Albrecht}, \citenamefont {Hunger}, \citenamefont {Doert}, \citenamefont {Vojta},\ and\ \citenamefont {B\"uchner}}]{Wolter2017}%
  \BibitemOpen
  \bibfield  {author} {\bibinfo {author} {\bibfnamefont {A.~U.~B.}\ \bibnamefont {Wolter}}, \bibinfo {author} {\bibfnamefont {L.~T.}\ \bibnamefont {Corredor}}, \bibinfo {author} {\bibfnamefont {L.}~\bibnamefont {Janssen}}, \bibinfo {author} {\bibfnamefont {K.}~\bibnamefont {Nenkov}}, \bibinfo {author} {\bibfnamefont {S.}~\bibnamefont {Sch\"onecker}}, \bibinfo {author} {\bibfnamefont {S.-H.}\ \bibnamefont {Do}}, \bibinfo {author} {\bibfnamefont {K.-Y.}\ \bibnamefont {Choi}}, \bibinfo {author} {\bibfnamefont {R.}~\bibnamefont {Albrecht}}, \bibinfo {author} {\bibfnamefont {J.}~\bibnamefont {Hunger}}, \bibinfo {author} {\bibfnamefont {T.}~\bibnamefont {Doert}}, \bibinfo {author} {\bibfnamefont {M.}~\bibnamefont {Vojta}},\ and\ \bibinfo {author} {\bibfnamefont {B.}~\bibnamefont {B\"uchner}},\ }\bibfield  {title} {\bibinfo {title} {{Field-induced quantum criticality in the Kitaev system $\alpha$-{RuCl}$_{3}$}},\ }\href {https://doi.org/10.1103/PhysRevB.96.041405} {\bibfield  {journal} {\bibinfo  {journal} {Phys. Rev. B}\ }\textbf {\bibinfo {volume} {96}},\ \bibinfo {pages} {041405(R)} (\bibinfo {year} {2017})}\BibitemShut {NoStop}%
\bibitem [{\citenamefont {Motome}\ and\ \citenamefont {Nasu}(2020)}]{Motome2020}%
  \BibitemOpen
  \bibfield  {author} {\bibinfo {author} {\bibfnamefont {Y.}~\bibnamefont {Motome}}\ and\ \bibinfo {author} {\bibfnamefont {J.}~\bibnamefont {Nasu}},\ }\bibfield  {title} {\bibinfo {title} {Hunting {Majorana} fermions in {Kitaev} magnets},\ }\href {https://doi.org/10.7566/JPSJ.89.012002} {\bibfield  {journal} {\bibinfo  {journal} {J. Phys. Soc. Jpn.}\ }\textbf {\bibinfo {volume} {89}},\ \bibinfo {pages} {012002} (\bibinfo {year} {2020})}\BibitemShut {NoStop}%
\bibitem [{\citenamefont {Do}\ \emph {et~al.}(2017)\citenamefont {Do}, \citenamefont {Park}, \citenamefont {Yoshitake}, \citenamefont {Nasu}, \citenamefont {Motome}, \citenamefont {Kwon}, \citenamefont {Adroja}, \citenamefont {Voneshen}, \citenamefont {Kim}, \citenamefont {Jang}, \citenamefont {Park}, \citenamefont {Choi},\ and\ \citenamefont {Ji}}]{Do2017}%
  \BibitemOpen
  \bibfield  {author} {\bibinfo {author} {\bibfnamefont {S.-H.}\ \bibnamefont {Do}}, \bibinfo {author} {\bibfnamefont {S.-Y.}\ \bibnamefont {Park}}, \bibinfo {author} {\bibfnamefont {J.}~\bibnamefont {Yoshitake}}, \bibinfo {author} {\bibfnamefont {J.}~\bibnamefont {Nasu}}, \bibinfo {author} {\bibfnamefont {Y.}~\bibnamefont {Motome}}, \bibinfo {author} {\bibfnamefont {Y.~S.}\ \bibnamefont {Kwon}}, \bibinfo {author} {\bibfnamefont {D.~T.}\ \bibnamefont {Adroja}}, \bibinfo {author} {\bibfnamefont {D.~J.}\ \bibnamefont {Voneshen}}, \bibinfo {author} {\bibfnamefont {K.}~\bibnamefont {Kim}}, \bibinfo {author} {\bibfnamefont {T.-H.}\ \bibnamefont {Jang}}, \bibinfo {author} {\bibfnamefont {J.-H.}\ \bibnamefont {Park}}, \bibinfo {author} {\bibfnamefont {K.-Y.}\ \bibnamefont {Choi}},\ and\ \bibinfo {author} {\bibfnamefont {S.}~\bibnamefont {Ji}},\ }\bibfield  {title} {\bibinfo {title} {{Majorana} fermions in the {Kitaev} quantum spin system $\alpha$-{RuCl}$_3$},\ }\href {https://doi.org/10.1038/nphys4264} {\bibfield  {journal} {\bibinfo  {journal} {Nat. Phys.}\ }\textbf {\bibinfo {volume} {13}},\ \bibinfo {pages} {1079} (\bibinfo {year} {2017})}\BibitemShut {NoStop}%
\bibitem [{\citenamefont {Suetsugu}\ \emph {et~al.}(2022)\citenamefont {Suetsugu}, \citenamefont {Ukai}, \citenamefont {Shimomura}, \citenamefont {Kamimura}, \citenamefont {Asaba}, \citenamefont {Kasahara}, \citenamefont {Kurita}, \citenamefont {Tanaka}, \citenamefont {Shibauchi}, \citenamefont {Nasu}, \citenamefont {Motome},\ and\ \citenamefont {Matsuda}}]{Suetsugu2022}%
  \BibitemOpen
  \bibfield  {author} {\bibinfo {author} {\bibfnamefont {S.}~\bibnamefont {Suetsugu}}, \bibinfo {author} {\bibfnamefont {Y.}~\bibnamefont {Ukai}}, \bibinfo {author} {\bibfnamefont {M.}~\bibnamefont {Shimomura}}, \bibinfo {author} {\bibfnamefont {M.}~\bibnamefont {Kamimura}}, \bibinfo {author} {\bibfnamefont {T.}~\bibnamefont {Asaba}}, \bibinfo {author} {\bibfnamefont {Y.}~\bibnamefont {Kasahara}}, \bibinfo {author} {\bibfnamefont {N.}~\bibnamefont {Kurita}}, \bibinfo {author} {\bibfnamefont {H.}~\bibnamefont {Tanaka}}, \bibinfo {author} {\bibfnamefont {T.}~\bibnamefont {Shibauchi}}, \bibinfo {author} {\bibfnamefont {J.}~\bibnamefont {Nasu}}, \bibinfo {author} {\bibfnamefont {Y.}~\bibnamefont {Motome}},\ and\ \bibinfo {author} {\bibfnamefont {Y.}~\bibnamefont {Matsuda}},\ }\bibfield  {title} {\bibinfo {title} {Evidence for a phase transition in the quantum spin liquid state of a {Kitaev} candidate {$\alpha$-RuCl$_3$}},\ }\href {https://doi.org/10.7566/JPSJ.91.124703} {\bibfield  {journal} {\bibinfo  {journal} {J. Phys. Soc. Jpn.}\ }\textbf {\bibinfo {volume} {91}},\ \bibinfo {pages} {124703} (\bibinfo {year} {2022})}\BibitemShut {NoStop}%
\bibitem [{\citenamefont {Kimchi}\ \emph {et~al.}(2018)\citenamefont {Kimchi}, \citenamefont {Sheckelton}, \citenamefont {McQueen},\ and\ \citenamefont {Lee}}]{Kimchi2018}%
  \BibitemOpen
  \bibfield  {author} {\bibinfo {author} {\bibfnamefont {I.}~\bibnamefont {Kimchi}}, \bibinfo {author} {\bibfnamefont {J.~P.}\ \bibnamefont {Sheckelton}}, \bibinfo {author} {\bibfnamefont {T.~M.}\ \bibnamefont {McQueen}},\ and\ \bibinfo {author} {\bibfnamefont {P.~A.}\ \bibnamefont {Lee}},\ }\bibfield  {title} {\bibinfo {title} {{Scaling and data collapse from local moments in frustrated disordered quantum spin systems}},\ }\href {https://doi.org/10.1038/s41467-018-06800-2} {\bibfield  {journal} {\bibinfo  {journal} {Nat. Commun.}\ }\textbf {\bibinfo {volume} {9}},\ \bibinfo {pages} {4367} (\bibinfo {year} {2018})}\BibitemShut {NoStop}%
\bibitem [{\citenamefont {Murayama}\ \emph {et~al.}(2020)\citenamefont {Murayama}, \citenamefont {Sato}, \citenamefont {Taniguchi}, \citenamefont {Kurihara}, \citenamefont {Xing}, \citenamefont {Huang}, \citenamefont {Kasahara}, \citenamefont {Kasahara}, \citenamefont {Kimchi}, \citenamefont {Yoshida}, \citenamefont {Iwasa}, \citenamefont {Mizukami}, \citenamefont {Shibauchi}, \citenamefont {Konczykowski},\ and\ \citenamefont {Matsuda}}]{Murayama2020}%
  \BibitemOpen
  \bibfield  {author} {\bibinfo {author} {\bibfnamefont {H.}~\bibnamefont {Murayama}}, \bibinfo {author} {\bibfnamefont {Y.}~\bibnamefont {Sato}}, \bibinfo {author} {\bibfnamefont {T.}~\bibnamefont {Taniguchi}}, \bibinfo {author} {\bibfnamefont {R.}~\bibnamefont {Kurihara}}, \bibinfo {author} {\bibfnamefont {X.~Z.}\ \bibnamefont {Xing}}, \bibinfo {author} {\bibfnamefont {W.}~\bibnamefont {Huang}}, \bibinfo {author} {\bibfnamefont {S.}~\bibnamefont {Kasahara}}, \bibinfo {author} {\bibfnamefont {Y.}~\bibnamefont {Kasahara}}, \bibinfo {author} {\bibfnamefont {I.}~\bibnamefont {Kimchi}}, \bibinfo {author} {\bibfnamefont {M.}~\bibnamefont {Yoshida}}, \bibinfo {author} {\bibfnamefont {Y.}~\bibnamefont {Iwasa}}, \bibinfo {author} {\bibfnamefont {Y.}~\bibnamefont {Mizukami}}, \bibinfo {author} {\bibfnamefont {T.}~\bibnamefont {Shibauchi}}, \bibinfo {author} {\bibfnamefont {M.}~\bibnamefont {Konczykowski}},\ and\ \bibinfo {author} {\bibfnamefont {Y.}~\bibnamefont {Matsuda}},\ }\bibfield  {title} {\bibinfo {title} {{Effect of quenched disorder on the quantum spin liquid state of the triangular-lattice antiferromagnet $1T$-TaS$_{2}$}},\ }\href {https://doi.org/10.1103/PhysRevResearch.2.013099} {\bibfield  {journal} {\bibinfo  {journal} {Phys. Rev. Res.}\ }\textbf {\bibinfo {volume} {2}},\ \bibinfo {pages} {013099} (\bibinfo {year} {2020})}\BibitemShut {NoStop}%
\bibitem [{\citenamefont {Murayama}\ \emph {et~al.}(2022)\citenamefont {Murayama}, \citenamefont {Tominaga}, \citenamefont {Asaba}, \citenamefont {Silva}, \citenamefont {Sato}, \citenamefont {Suzuki}, \citenamefont {Ukai}, \citenamefont {Suetsugu}, \citenamefont {Kasahara}, \citenamefont {Okuma}, \citenamefont {Kimchi},\ and\ \citenamefont {Matsuda}}]{Murayama2022}%
  \BibitemOpen
  \bibfield  {author} {\bibinfo {author} {\bibfnamefont {H.}~\bibnamefont {Murayama}}, \bibinfo {author} {\bibfnamefont {T.}~\bibnamefont {Tominaga}}, \bibinfo {author} {\bibfnamefont {T.}~\bibnamefont {Asaba}}, \bibinfo {author} {\bibfnamefont {A.~d.}\ \bibnamefont {Silva}}, \bibinfo {author} {\bibfnamefont {Y.}~\bibnamefont {Sato}}, \bibinfo {author} {\bibfnamefont {H.}~\bibnamefont {Suzuki}}, \bibinfo {author} {\bibfnamefont {Y.}~\bibnamefont {Ukai}}, \bibinfo {author} {\bibfnamefont {S.}~\bibnamefont {Suetsugu}}, \bibinfo {author} {\bibfnamefont {Y.}~\bibnamefont {Kasahara}}, \bibinfo {author} {\bibfnamefont {R.}~\bibnamefont {Okuma}}, \bibinfo {author} {\bibfnamefont {I.}~\bibnamefont {Kimchi}},\ and\ \bibinfo {author} {\bibfnamefont {Y.}~\bibnamefont {Matsuda}},\ }\bibfield  {title} {\bibinfo {title} {{Universal scaling of specific heat in the $S=\frac{1}{2}$ quantum kagome antiferromagnet herbertsmithite}},\ }\href {https://doi.org/10.1103/PhysRevB.106.174406} {\bibfield  {journal} {\bibinfo  {journal} {Phys. Rev. B}\ }\textbf {\bibinfo {volume} {106}},\ \bibinfo {pages} {174406} (\bibinfo {year} {2022})}\BibitemShut {NoStop}%
\bibitem [{\citenamefont {Altland}\ and\ \citenamefont {Zirnbauer}(1997)}]{Altland1997}%
  \BibitemOpen
  \bibfield  {author} {\bibinfo {author} {\bibfnamefont {A.}~\bibnamefont {Altland}}\ and\ \bibinfo {author} {\bibfnamefont {M.~R.}\ \bibnamefont {Zirnbauer}},\ }\bibfield  {title} {\bibinfo {title} {{Nonstandard symmetry classes in mesoscopic normal-superconducting hybrid structures}},\ }\href {https://doi.org/10.1103/PhysRevB.55.1142} {\bibfield  {journal} {\bibinfo  {journal} {Phys. Rev. B}\ }\textbf {\bibinfo {volume} {55}},\ \bibinfo {pages} {1142} (\bibinfo {year} {1997})}\BibitemShut {NoStop}%
\bibitem [{\citenamefont {O'Brien}\ \emph {et~al.}(2016)\citenamefont {O'Brien}, \citenamefont {Hermanns},\ and\ \citenamefont {Trebst}}]{O'Brien2016}%
  \BibitemOpen
  \bibfield  {author} {\bibinfo {author} {\bibfnamefont {K.}~\bibnamefont {O'Brien}}, \bibinfo {author} {\bibfnamefont {M.}~\bibnamefont {Hermanns}},\ and\ \bibinfo {author} {\bibfnamefont {S.}~\bibnamefont {Trebst}},\ }\bibfield  {title} {\bibinfo {title} {{Classification of gapless $Z_{2}$ spin liquids in three-dimensional Kitaev models}},\ }\href {https://doi.org/10.1103/PhysRevB.93.085101} {\bibfield  {journal} {\bibinfo  {journal} {Phys. Rev. B}\ }\textbf {\bibinfo {volume} {93}},\ \bibinfo {pages} {085101} (\bibinfo {year} {2016})}\BibitemShut {NoStop}%
\bibitem [{\citenamefont {Ryu}\ and\ \citenamefont {Hatsugai}(2001)}]{Ryu2001}%
  \BibitemOpen
  \bibfield  {author} {\bibinfo {author} {\bibfnamefont {S.}~\bibnamefont {Ryu}}\ and\ \bibinfo {author} {\bibfnamefont {Y.}~\bibnamefont {Hatsugai}},\ }\bibfield  {title} {\bibinfo {title} {{Singular density of states of disordered Dirac fermions in chiral models}},\ }\href {https://doi.org/10.1103/PhysRevB.65.033301} {\bibfield  {journal} {\bibinfo  {journal} {Phys. Rev. B}\ }\textbf {\bibinfo {volume} {65}},\ \bibinfo {pages} {033301} (\bibinfo {year} {2001})}\BibitemShut {NoStop}%
\bibitem [{\citenamefont {Prozorov}\ \emph {et~al.}(2014)\citenamefont {Prozorov}, \citenamefont {Ko{\' n}czykowski}, \citenamefont {Tanatar}, \citenamefont {Thaler}, \citenamefont {Bud'ko}, \citenamefont {Canfield}, \citenamefont {Mishra},\ and\ \citenamefont {Hirschfeld}}]{Prozorov2014}%
  \BibitemOpen
  \bibfield  {author} {\bibinfo {author} {\bibfnamefont {R.}~\bibnamefont {Prozorov}}, \bibinfo {author} {\bibfnamefont {M.}~\bibnamefont {Ko{\' n}czykowski}}, \bibinfo {author} {\bibfnamefont {M.~A.}\ \bibnamefont {Tanatar}}, \bibinfo {author} {\bibfnamefont {A.}~\bibnamefont {Thaler}}, \bibinfo {author} {\bibfnamefont {S.~L.}\ \bibnamefont {Bud'ko}}, \bibinfo {author} {\bibfnamefont {P.~C.}\ \bibnamefont {Canfield}}, \bibinfo {author} {\bibfnamefont {V.}~\bibnamefont {Mishra}},\ and\ \bibinfo {author} {\bibfnamefont {P.~J.}\ \bibnamefont {Hirschfeld}},\ }\bibfield  {title} {\bibinfo {title} {{Effect of Electron Irradiation on Superconductivity in Single Crystals of Ba(Fe$_{1-x}$Ru$_{x}$)$_{2}$As$_{2}$ ($x=0.24$)}},\ }\href {https://doi.org/10.1103/PhysRevX.4.041032} {\bibfield  {journal} {\bibinfo  {journal} {Phys. Rev. X}\ }\textbf {\bibinfo {volume} {4}},\ \bibinfo {pages} {041032} (\bibinfo {year} {2014})}\BibitemShut {NoStop}%
\bibitem [{\citenamefont {Bois}(2011)}]{Bois1987}%
  \BibitemOpen
  \bibfield  {author} {\bibinfo {author} {\bibfnamefont {P.}~\bibnamefont {Bois}},\ }\href {https://inis.iaea.org/collection/NCLCollectionStore/_Public/18/082/18082185.pdf} {\emph {\bibinfo {title} {Etude des d{\' e}fauts Ponctuels dans le bismuth.}}},\ CEA report\ (\bibinfo {year} {2011})\BibitemShut {NoStop}%
\bibitem [{\citenamefont {Roppongi}\ \emph {et~al.}(2023)\citenamefont {Roppongi}, \citenamefont {Ishihara}, \citenamefont {Tanaka}, \citenamefont {Ogawa}, \citenamefont {Okada}, \citenamefont {Liu}, \citenamefont {Mukasa}, \citenamefont {Mizukami}, \citenamefont {Uwatoko}, \citenamefont {Grasset}, \citenamefont {Konczykowski}, \citenamefont {Ortiz}, \citenamefont {Wilson}, \citenamefont {Hashimoto},\ and\ \citenamefont {Shibauchi}}]{Roppongi2023}%
  \BibitemOpen
  \bibfield  {author} {\bibinfo {author} {\bibfnamefont {M.}~\bibnamefont {Roppongi}}, \bibinfo {author} {\bibfnamefont {K.}~\bibnamefont {Ishihara}}, \bibinfo {author} {\bibfnamefont {Y.}~\bibnamefont {Tanaka}}, \bibinfo {author} {\bibfnamefont {K.}~\bibnamefont {Ogawa}}, \bibinfo {author} {\bibfnamefont {K.}~\bibnamefont {Okada}}, \bibinfo {author} {\bibfnamefont {S.}~\bibnamefont {Liu}}, \bibinfo {author} {\bibfnamefont {K.}~\bibnamefont {Mukasa}}, \bibinfo {author} {\bibfnamefont {Y.}~\bibnamefont {Mizukami}}, \bibinfo {author} {\bibfnamefont {Y.}~\bibnamefont {Uwatoko}}, \bibinfo {author} {\bibfnamefont {R.}~\bibnamefont {Grasset}}, \bibinfo {author} {\bibfnamefont {M.}~\bibnamefont {Konczykowski}}, \bibinfo {author} {\bibfnamefont {B.~R.}\ \bibnamefont {Ortiz}}, \bibinfo {author} {\bibfnamefont {S.~D.}\ \bibnamefont {Wilson}}, \bibinfo {author} {\bibfnamefont {K.}~\bibnamefont {Hashimoto}},\ and\ \bibinfo {author} {\bibfnamefont {T.}~\bibnamefont {Shibauchi}},\ }\bibfield  {title} {\bibinfo {title} {{Bulk evidence of anisotropic $s$-wave pairing with no sign change in the kagome superconductor CsV$_3$Sb$_5$}},\ }\href {https://doi.org/10.1038/s41467-023-36273-x} {\bibfield  {journal} {\bibinfo  {journal} {Nat. Commun.}\ }\textbf {\bibinfo {volume} {14}},\ \bibinfo {pages} {667} (\bibinfo {year} {2023})}\BibitemShut {NoStop}%
\bibitem [{\citenamefont {Nagai}\ \emph {et~al.}(2020)\citenamefont {Nagai}, \citenamefont {Jinno}, \citenamefont {Yoshitake}, \citenamefont {Nasu}, \citenamefont {Motome}, \citenamefont {Itoh},\ and\ \citenamefont {Shimizu}}]{Nagai2020}%
  \BibitemOpen
  \bibfield  {author} {\bibinfo {author} {\bibfnamefont {Y.}~\bibnamefont {Nagai}}, \bibinfo {author} {\bibfnamefont {T.}~\bibnamefont {Jinno}}, \bibinfo {author} {\bibfnamefont {J.}~\bibnamefont {Yoshitake}}, \bibinfo {author} {\bibfnamefont {J.}~\bibnamefont {Nasu}}, \bibinfo {author} {\bibfnamefont {Y.}~\bibnamefont {Motome}}, \bibinfo {author} {\bibfnamefont {M.}~\bibnamefont {Itoh}},\ and\ \bibinfo {author} {\bibfnamefont {Y.}~\bibnamefont {Shimizu}},\ }\bibfield  {title} {\bibinfo {title} {Two-step gap opening across the quantum critical point in the {Kitaev} honeycomb magnet $\alpha$-{RuCl}$_{3}$},\ }\href {https://doi.org/10.1103/PhysRevB.101.020414} {\bibfield  {journal} {\bibinfo  {journal} {Phys. Rev. B}\ }\textbf {\bibinfo {volume} {101}},\ \bibinfo {pages} {020414} (\bibinfo {year} {2020})}\BibitemShut {NoStop}%
\bibitem [{\citenamefont {Kim}\ \emph {et~al.}(2023)\citenamefont {Kim}, \citenamefont {Horsley}, \citenamefont {Ruff}, \citenamefont {Moreno},\ and\ \citenamefont {Kim}}]{Kim2023}%
  \BibitemOpen
  \bibfield  {author} {\bibinfo {author} {\bibfnamefont {S.}~\bibnamefont {Kim}}, \bibinfo {author} {\bibfnamefont {E.}~\bibnamefont {Horsley}}, \bibinfo {author} {\bibfnamefont {J.~P.~C.}\ \bibnamefont {Ruff}}, \bibinfo {author} {\bibfnamefont {B.~D.}\ \bibnamefont {Moreno}},\ and\ \bibinfo {author} {\bibfnamefont {Y.-J.}\ \bibnamefont {Kim}},\ }\href@noop {} {\bibinfo {title} {Structural transition and magnetic anisotropy in {$\alpha$-RuCl$_{3}$}}} (\bibinfo {year} {2023}),\ \Eprint {https://arxiv.org/abs/2311.04000} {arXiv:2311.04000 [cond-mat.str-el]} \BibitemShut {NoStop}%
\end{thebibliography}%

\end{document}